\documentclass{aa}
\usepackage[varg]{txfonts}
\usepackage{graphicx}
\usepackage{ulem}
\usepackage{color}
\usepackage{amsmath} 
\usepackage{natbib}

\begin{document}

\title{Differential interferometric phases at high spectral resolution\\as a sensitive physical diagnostic of circumstellar disks}
\author{D.~M.~Faes\inst{\ref{inst1},\ref{inst5}}
\and A.~C.~Carciofi\inst{\ref{inst1}}
\and Th.~Rivinius\inst{\ref{inst2}} 
\and S.~\v{S}tefl\inst{\ref{inst3}} 
\and D.~Baade\inst{\ref{inst4}}
\and A.~Domiciano~de~Souza\inst{\ref{inst5}} 
}

\institute{{Instituto de Astronomia, Geof\'isica e Ci\^encias Atmosf\'ericas, Universidade de S\~ao Paulo, Rua do Mat\~ao 1226, Cidade Universit\'aria, 05508-900, S\~ao Paulo, SP, Brazil, moser@usp.br, carciofi@usp.br}\label{inst1}
\and{European Organisation for Astronomical Research in the Southern Hemisphere, Casilla 19001, Santiago 19, Chile}\label{inst2}
\and{European Organisation for Astronomical Research in the Southern Hemisphere / ALMA,
Alonso de Córdova 3107, Santiago, Chile}\label{inst3}
\and{European Organisation for Astronomical Research in the Southern Hemisphere, Karl-Schwarzschild-Str. 2, 85748 Garching bei M\"unchen, Germany}\label{inst4}
\and{Lab. J.-L. Lagrange, UMR 7293 - Observatoire de la C\^{o}te d'Azur (OCA), Univ. de Nice-Sophia
Antipolis (UNS), CNRS, Valrose,  06108 Nice, France}\label{inst5}
}

\date{Received 18 February 2013 / 
         Accepted DD MMMMM YYYY}
\abstract
{ The circumstellar disks ejected by many rapidly rotating B stars (so-called Be stars) offer the rare opportunity of studying the structure and dynamics of gaseous disks at high spectral as well as angular resolution.}
{ This paper explores a newly identified effect in spectro-interferometric phase that can be used for probing the inner regions of gaseous edge-on disks on a scale of a few stellar radii. }
{ The origin of this
effect (dubbed central quasi-emission phase signature, CQE-PS) lies in the velocity-dependent line absorption {of photospheric radiation} by the {circumstellar} disk.  At high spectral and marginal interferometric resolution, photocenter displacements between star and isovelocity regions in the Keplerian disk reveal themselves through small interferometric phase shifts.  To investigate the diagnostic potential of this effect, a series of models are presented, based on detailed radiative transfer calculations in a viscous decretion disk.}
{ Amplitude and detailed shape of the CQE-PS depend sensitively on disk density and size and on the radial distribution of the material with characteristic shapes in differential phase diagrams.  In addition, useful lower limits to the angular size of the central stars can be derived even when the system is almost unresolved.}
{  The full power of this diagnostic tool can be expected if it can be applied to observations over a  full life-cycle of a disk from first ejection through final dispersal, over a full cycle of disk oscillations, or over a full orbital period in a binary system. 
}


\titlerunning{Differential interferometric phases at high spectral resolution as a physical diagnostic of circumstellar disks}
\maketitle

\section{Introduction} \label{intro}

Circumstellar (CS) disks can assume importance at various phases of the life cycles of single and binary stars.
They are the place of fundamental physical processes related to properties and evolution of their central objects.  Be stars are a unique class of objects that allow studying the creation, evolution and destruction of CS disks. In these main-sequence objects, the disks are fed from mass lost from their rapidly rotating central stars \citep{por03}. Questions such as how the CS material can be supported against the stellar gravity and how angular momentum can be transferred from inner to external regions remained open for a long time. In the past two decades, optical and infrared interferometry played a key role in developing our understanding of these astrophysical systems. For instance, the pioneering measurements of \citet{qui97} provided strong evidence that the CS matter around Be stars is distributed in a geometrically thin structure (i.e., a disk). More recently, detailed measurements have demonstrated that Be disks rotate in a Keplerian fashion \citep{mei07,kra11,whe12}.

The strong developments in the observations were accompanied by significant progress in our theoretical understanding of these systems. Recently, a consensus is emerging that the viscous decretion disk (VDD) model, first suggested by \cite{lee91} and subsequently developed by \citet{bjo97}, and \citet{oka01}, among others, is the best candidate to explain the observed properties of Be disks \citep{car11,mcg11}.
{Detailed hydrodynamical calculations \citep[e.g.,][]{oka01,jon08} have demonstrated that VDDs rotate in a nearly Keplerian fashion, in agreement with the results from spectro-interferometry \citep{mei07,kra11}.}

Before the spectro-interferometric era, one of the first strong indications that Be disks are Keplerian came from high-resolution spectroscopy ($R>30000$) of shell stars, {which are stars with strongly rotationally broadened photospheric lines and additional narrow absorption lines. They are understood to be ordinary Be stars seen edge-on, so that the line-of-sight toward the star probes the CS  equatorial disk}.
\citet{han95} and \citet{riv99} studied the so-called \textit{central quasi-emission} (hereafter, CQE) peaks, where the disk, under certain circumstances, causes a cusp in the deepest region of the line profile of shell stars. The existence of the CQE implies slow radial motions of the gas (much slower than the sound speed), which means that the disk is supported by rotation.

The same mechanism that produces the CQE
observed in shell-star line profiles can cause important changes in the
monochromatic intensity maps of the Be star plus disk system, with observable effects on the
interferometric quantities. 
This effect is dubbed \textit{CQE phase signature} \citep{fae12}, and here its diagnostic potential is investigated.

Similar to the spectroscopic CQEs, the CQE Phase Signature (hereafter, CQE-PS) can only be
studied by high {spectral} resolution observations ({$R>2500$}), such as can be obtained {by the new generation of interferometers (AMBER/VLTI at near-infrared, \citealt{pet07}, and VEGA/CHARA at visible range, \citealt{mou09})}. 

{Part of the variability observed in Be stars is associated with
structural changes in their CS disks. The temporal
evolution of the observables can only be put in some sort of conceptual frame with the
understanding of the physical processes that control the CS dynamics.
\citet{car12} studied
the $V$-band magnitude evolution of 28 CMa during a phase of disk dissipation using the VDD model.
This allowed the determination of the
viscosity parameter governing the disk dynamics, as well its
corresponding mass decretion rate.
\citet{hau12} provided another example of how CS disk and stellar parameters
are intrinsically connected and how they can be inferred through a
follow-up of various observables. In particular, the temporal
evolution of the disk density (and corresponding continuum emission)
is related to the mass-loss rate and episodes of mass injection and
quiescence.}

{The initial studies made on the CQE profile {\citep{han95,riv99}} demonstrated its strong dependence on the distribution and velocity field of the CS material. Likewise, we show here that CQE-PS has a high sensitivity to structural components of the CS disk - and even to the stellar size. These parameters can then be related to the time evolution of the observables and other fundamental parameters of the system as in the cases mentioned above.}

\section{Model description} \label{refcase}

The mechanism behind the CQE-PS is the differential absorption of photospheric radiation by the CS disk {(Sect.~\ref{prfeat})}. This mechanism can be directly traced by the interferometric differential phase, and must be separated from other mechanisms that affect this quantity.

To study, characterize, and illustrate the CQE-PS, we adopted a realistic model for the Be + disk system. 
For the central star we adopted a rotationally deformed and gravity-darkened star whose parameters are typical of a B1\,Ve star (Table~\ref{tab:bemod}). {The stellar geometry is described by {an oblate} spheroid with stellar flux determined by the traditional {von Zeipel} effect \citep{von24}, where the local effective temperature is proportional to the local surface gravity, $T_\textrm{eff}\propto {g_\textrm{eff}}^{0.25}$.}
For the CS disk description, we adopted the VDD model.  For isothermal
viscous diffusion in the steady-state regime, a state reached after a
sufficiently long and stable decretion period \citep{hau12}, the disk volume density has a particularly simple form 
$n(r) = n_0 (R_{\rm eq}/r)^{-m}$, where $m=3.5$ \citep{bjo05}.

The interferometric quantities depend on the spatial resolution with which the object is seen. To explore different configurations we defined the quantity $\nu_{\rm obs}$, the ratio between 
the baseline length of the interferometer, {and the distance to the star}. The unit used is m\,pc$^{-1}$.

{The calculations presented in this paper were made with the three-dimensional, non-local radiative equilibrium code HDUST \citep{car06,car08}}. 

\begin{table}[!ht]
\begin{center}
\caption{Reference Be model parameters.}
{\small
\begin{tabular}{ccc}
\noalign{\smallskip}
\hline 
\hline 
\noalign{\smallskip}
Parameter & Symbol & Ref. Case \\
\noalign{\smallskip}
\hline 
\noalign{\smallskip}
Spectral type & - & B1\,V \\
Mass & $M$ & 11.0 $M_{\odot}$\\
Polar radius & $R_{\rm pole}$ & 4.9 $R_{\odot}$\\
Pole temperature & $T_{\rm pole}$ &  27440 K  \\
Luminosity & $L_{\star}$ & 10160 $L_{\odot}$ \\
Critical velocity & $v_{\rm crit}$ & 534.4 km/s \\
Rotation rate & $\Omega/\Omega_{\rm crit}$ & 0.8 \\
Oblateness & $R_{\rm eq}/R_{\rm pole}$ & 1.14 \\
Gravity darkening & $T_{\rm pole}/T_{\rm eq}$ & 1.16  \\ 
\noalign{\smallskip}
\hline 
\noalign{\smallskip}
Disk radius & $R_{\rm disk}$ & 10 $R_{\star}$  \\
Disk-density scale & $n_0$ & 10$^{13}\,\rm cm^{-3}$ \\
Density exponent & $m$ & 3.5 \\
\noalign{\smallskip}
\hline
\noalign{\smallskip}
Inclination angle & $i$ & 45$^\circ$  \\
Spectral resolving power & $R$ & 12\,000  \\
Baseline/distance & $\nu_{\rm obs}$ & 1 m pc$^{-1}$ \\
\noalign{\smallskip}
\hline
\end{tabular} }
\label{tab:bemod}
\end{center}	
\end{table}

\subsection{Interferometric phases in the marginally resolved case} \label{canon}

The quantity registered in interferometry is the normalized complex visibility $V(\textbf{u})$. It is related to the Fourier transform of the brightness distribution $I(\textbf{r}, \lambda)$ of a non-coherent and extended source on the plane of sky through the Van Cittert-Zernike theorem {\citep{bor80}}:
\begin{equation}
V(\textbf{u})=\frac{\iint I(\textbf{r}, \lambda) \exp{(-2\pi i \textbf{u}\cdot\textbf{r})} \,d^2\textbf{r}}{\iint I(\textbf{r}, \lambda) \,d^2\textbf{r}}=|V|\exp{(i\phi)} \,,
\label{eq:compvis}
\end{equation}
where $\lambda$ is the monochromatic wavelength and $\textbf{u}$ the spatial frequency of the projected baseline $\textbf{B}_\textrm{proj}/\lambda$. {$|V|$ ranges from 1 to 0 from a fully unresolved to a fully resolved object.}

{A common quantity used in high spectral resolution interferometry are the differential phases $\phi_\textrm{diff}$},
\begin{equation}
\phi_\textrm{diff}(\lambda,\lambda_\textrm{r})=\phi(\lambda)-\phi(\lambda_\textrm{r}) \,,
\label{eq:diffph}
\end{equation}
{where $\lambda_\textrm{r}$ is a wavelength of reference simultaneously observed.
One important characteristic of differential phases arises when the target is \textit{marginally resolved}, i.e., $\textbf{u}\cdot\textbf{r} \ll 1$. In this case, interferometric differential phases (hereafter DP or just phases) can directly map the target's photocenter} \citep[Eq. 4]{dom04}:
\begin{equation}
\phi_\textrm{diff}(\lambda, \lambda_r)=-2\pi\textbf{u}\cdot[\boldsymbol\epsilon(\lambda)-\boldsymbol\epsilon(\lambda_r)] \,;
\label{eq:photc}
\end{equation}
the $\boldsymbol\epsilon(\lambda)$ and $\boldsymbol\epsilon(\lambda_r)$ vectors are the \textit{photocenters} for $\lambda$ and $\lambda_r$. The usual procedure is to take $\lambda_r$ on the adjacent continuum of the spectral line, with $\boldsymbol\epsilon(\lambda_r)=0$. 

Photocenter measurements can be made at very high angular resolution using DP. One example is provided by \citet{ste09}, {who achieved <0.1-mas relative photocenter astrometry in the Be star $\zeta$ Tauri. }

We see below that the CQE-PS in DP occurs under quite specific circumstances, and its diagnostic potential is more relevant for some parameters than for others. To quantify this, we first studied a reference model for which the CQE-PS is not present, i.e., a model for which the canonical disk interferometric signatures apply. The adopted model parameters for the reference case are listed in the third column of Table~\ref{tab:bemod}.

\begin{figure*}[!t]
\centering
\includegraphics[width=.31\linewidth]{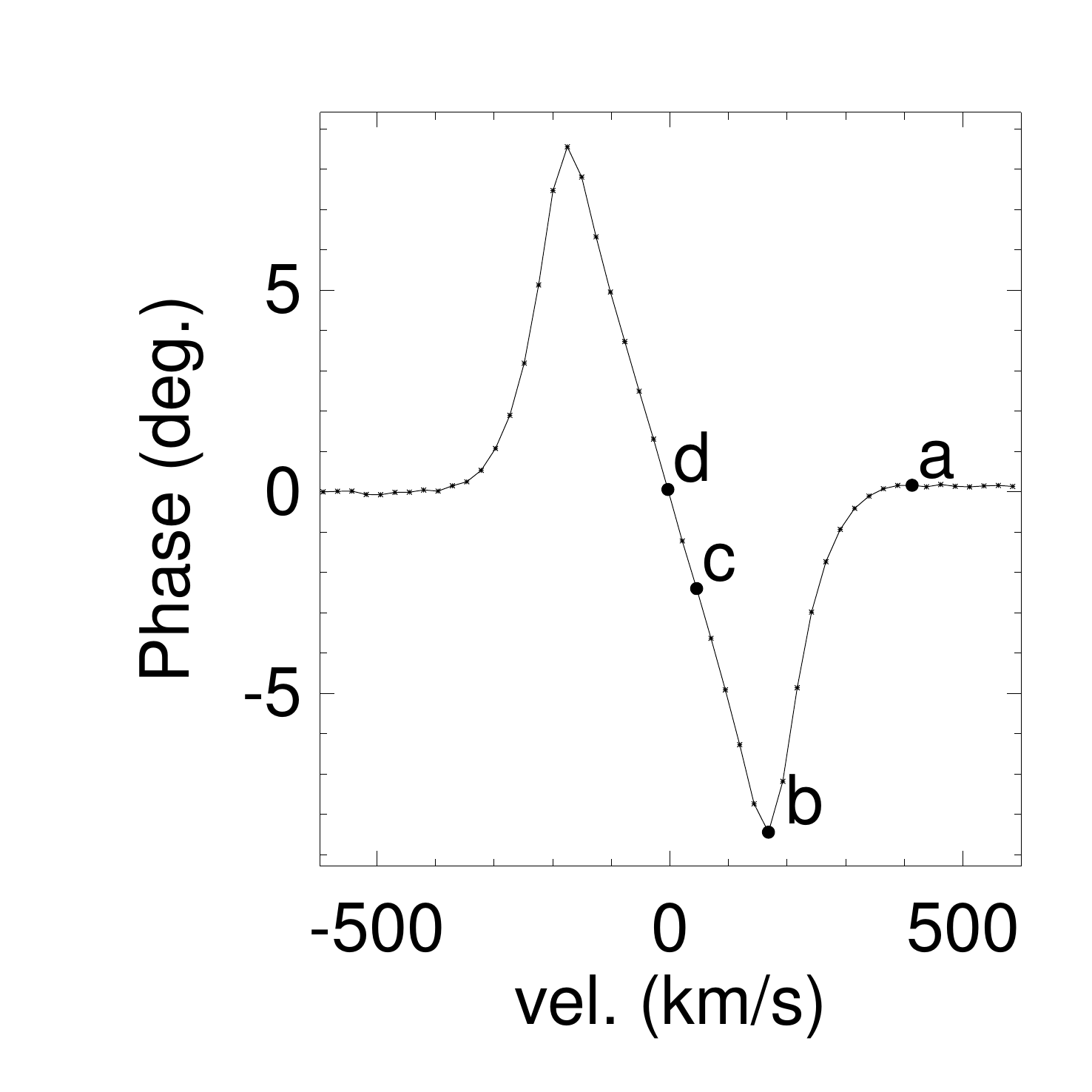}
\includegraphics[width=.62\linewidth]{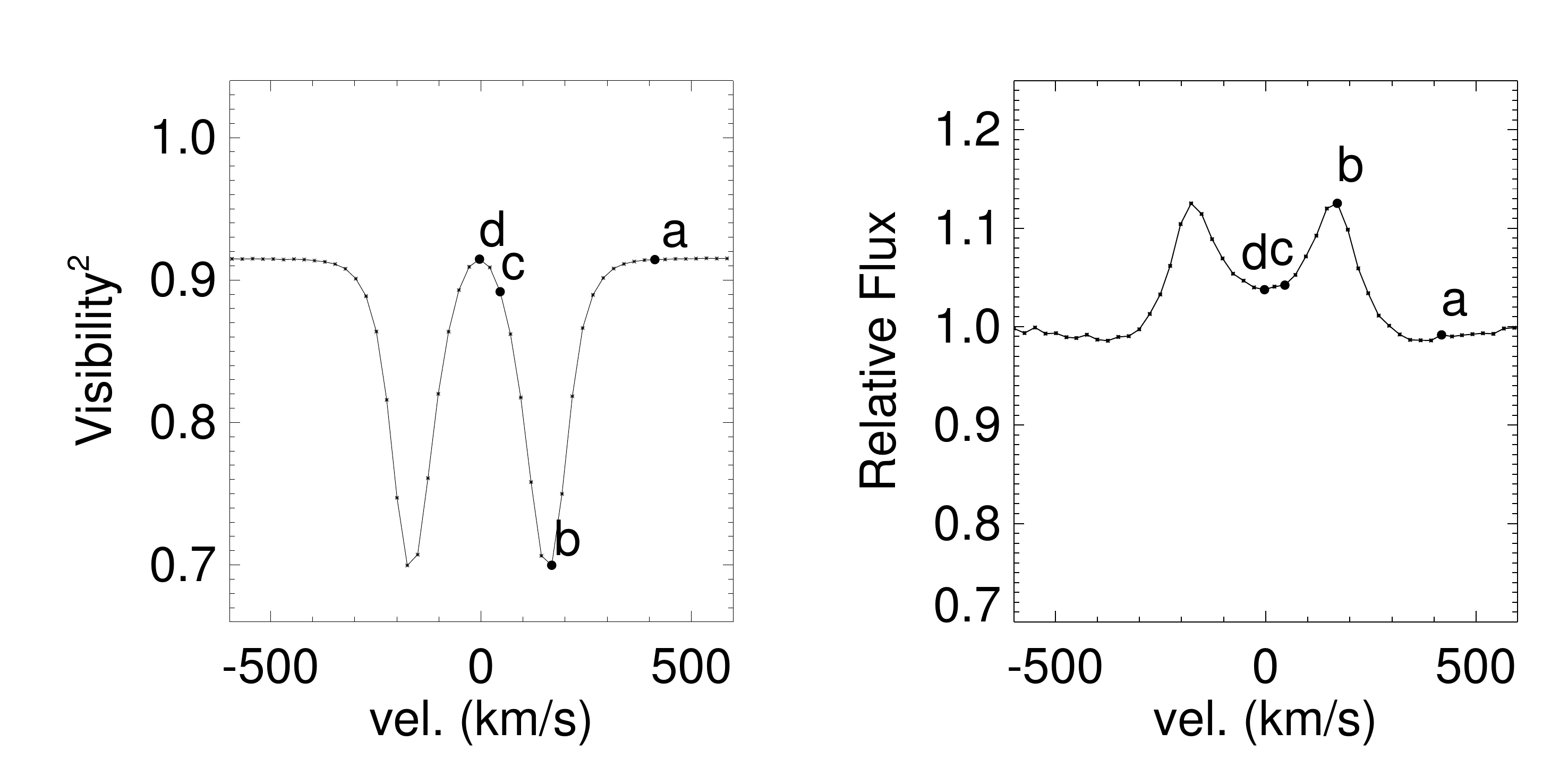} \\
\includegraphics[width=.9\linewidth]{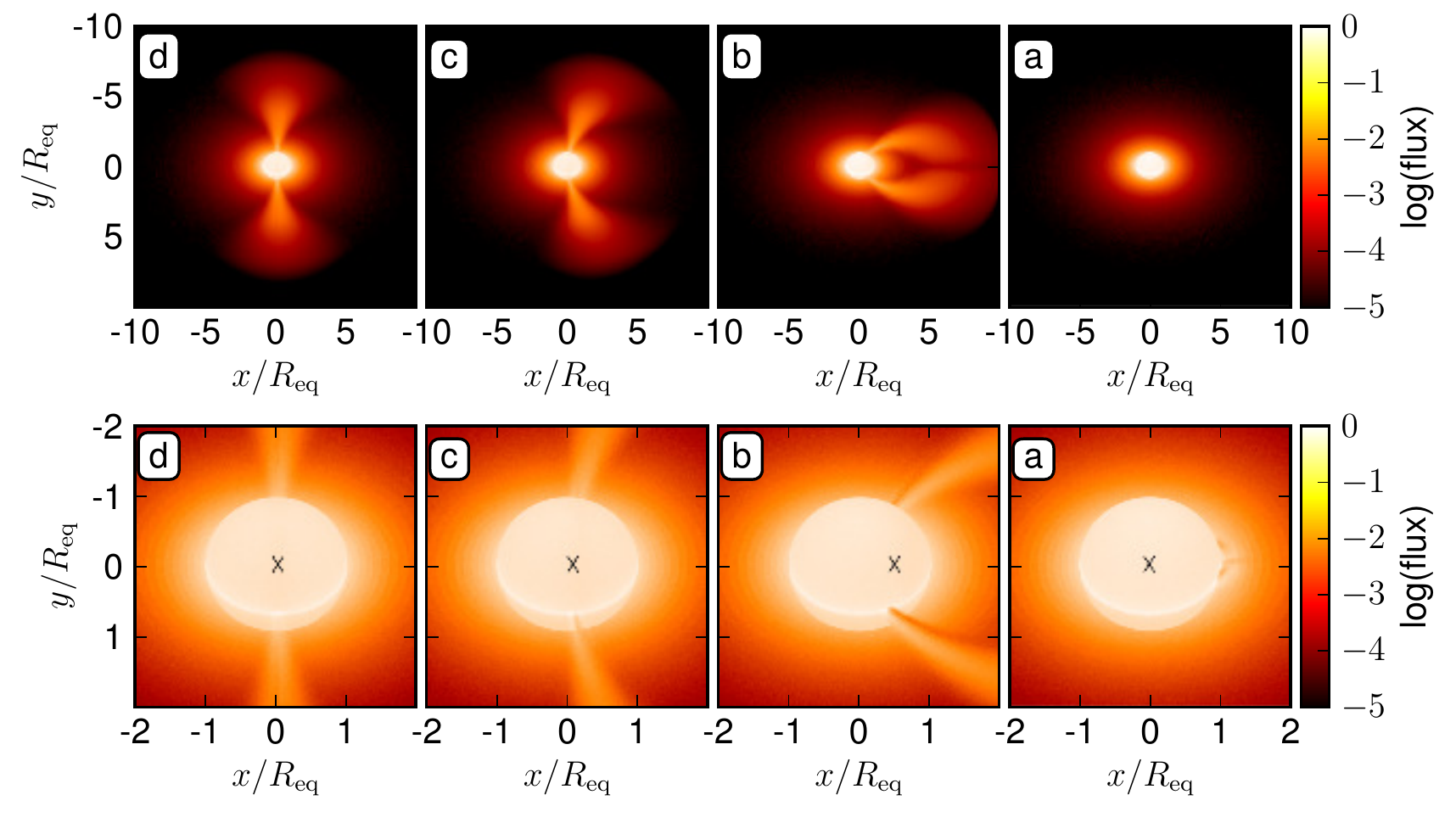} \\
\caption{\emph{Top 3 panels}: Interferometric Br$\gamma$ differential phases, visibilities, and line
  profile for the reference model.  Calculations
  were made for $\nu_{\rm obs}=1\,\rm m\,pc^{-1}$ {(target marginally resolved)}, and baseline orientation
  parallel to the disk equator. \emph{Bottom panels:} Model images for different spectral channels (as indicated in the upper
  panels) {at different spatial scales}. {The photocenter position is indicated by a black cross.} The disk inclination angle is $i=45^\circ$. Absorption bands can be identified at velocities (b) and (c) but they little alter the stellar flux in the line-of-sight.}
 \label{fig:classbe}
\end{figure*}

For rotating Be disks, it is {generally expected} that DP have a simple S-shaped profile \citep{ste96}.
This can be easily understood from Fig.~\ref{fig:classbe}, which shows the Br$\gamma$ line profile, the corresponding interferometric quantities and model images at different wavelengths for the reference model. Shown are results for $i=45^\circ$. At continuum wavelengths ({velocity} a, in Fig.~\ref{fig:classbe}) the images show the star surrounded by a centro-symmetric continuum emission. We defined the DP to be zero at those wavelengths.
As we move from the continuum toward line center (a to d), the line flux {at a given wavelength range} initially increases as a result of progressively larger emission lobes and then decreases as the emission area decreases toward line center (d). This creates the familiar W-shaped pattern in the visibilities and the S-shape pattern in the DPs.

\subsection{Interferometric phases in the resolved case} \label{nr_interf}

The association between the phase signal and the image's photocenter position is only possible if the marginally resolved condition is satisfied, since it permits the truncation of the Fourier expansion \citep{jan01}. This is particularly important for Be stars because their on-sky size is wavelength dependent, mainly across emission lines 
{whose monochromatic emission has a strong spatial dependence.}

\begin{figure}[!t]
\centering
\includegraphics[width=.7\linewidth]{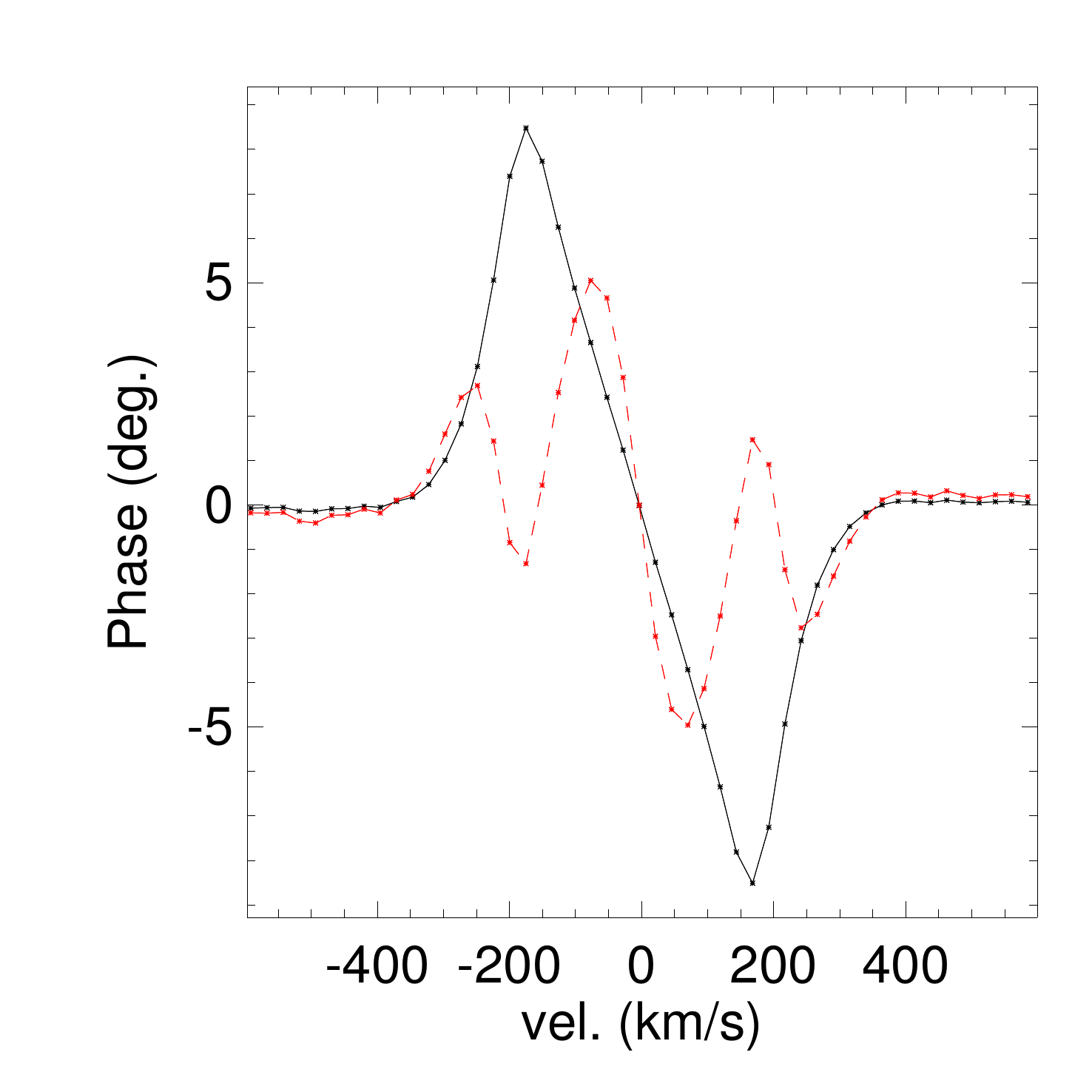}
\caption{Differential phase signal for the Be reference model ({{Sec.~\ref{refcase}}}). The black (full) line corresponds to $\nu_{\rm obs}= 1\,\rm m\,pc^{-1}$ (astrometric regime) and the  red (dashed) line to $\nu_{\rm obs}=2.5\,\rm m\,pc^{-1}$ (non-astrometric regime).}
 \label{fig:unres}
\end{figure}

\begin{figure}[!t]
\centering
\includegraphics[width=\linewidth]{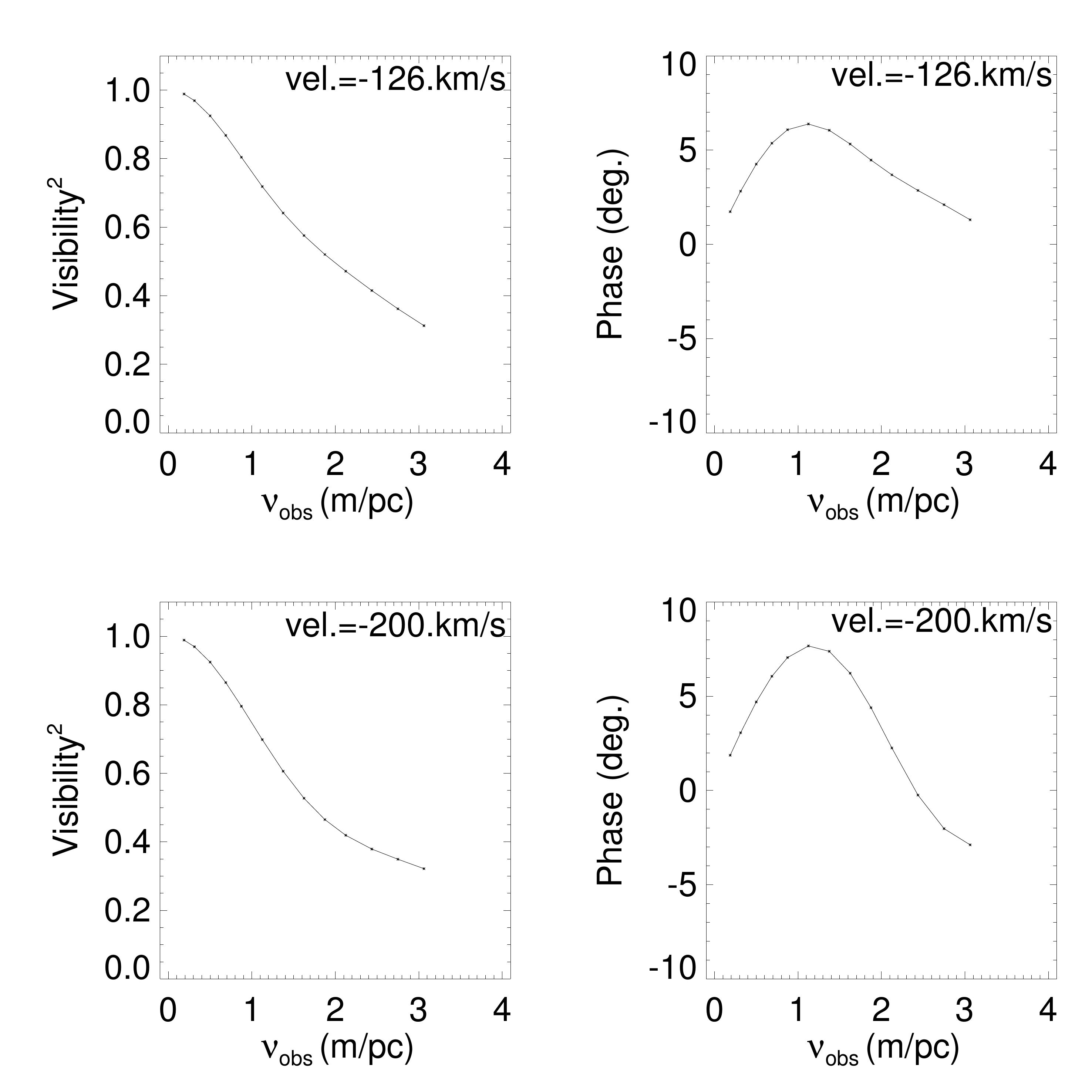}
\caption{Squared visibility and differential phases for the Be star reference model ({{Sec.~\ref{refcase}}}) at radial velocities of vel.$=-126\rm\,km\,s^{-1}$ and vel.$=-200\rm\,km\,s^{-1}$ {as a function of $\nu_{\rm obs}$, the ratio between the length of the baseline, and the distance of the target}. The linear {correspondence} of the phase with the increasing spatial resolution ($\nu_{\rm obs}\lesssim 1.5\rm\,m\,pc^{-1}$) characterizes the astrometric regime.}
 \label{fig:interf}
\end{figure}

Outside the astrometric regime (i.e., $\textbf{u}\cdot\textbf{r} \approx 1$ 
), Eq.~(\ref{eq:photc}) is no longer valid. A
complex phase behavior is then expected since higher-order terms of the
Fourier transform of the image are relevant
and a direct correspondence between the DP and the photocenter position does not hold any longer.
Fig.~\ref{fig:unres} compares the DP
signal of our reference model in the astrometric regime and outside it.  The
reason the curves are so different is explained in Fig.~\ref{fig:interf} that
plots the interferometric quantities for two wavelengths across Br$\gamma$ as
a function of $\nu_{\rm obs}${, the ratio between {|B$_{\rm proj}$|} and the target's distance  ($\nu_{\rm obs}\propto\textbf{u}\cdot\textbf{r} $). The astrometric regime corresponds to the linear growth of the phase shift with $\nu_{\rm obs}$ in Fig.~\ref{fig:interf} . When the target becomes marginally resolved ($\nu_{\rm obs} \gtrsim 1.5$), linear correspondence no longer holds, and in the fully resolved regime the phase shift \emph{may even decrease} with increasing $\nu_{\rm obs}$.}

The interferometric signature of a Be star comprises the stellar
  and the disk contributions; the visibility is equally affected by
  both, depending on the relative flux contributions, but the phase signal is dominated by
  the disk emission {since it comes from a large angular distance}. In general, the complex phase behavior seen in Fig~\ref{fig:unres}, including change of sign,
  occurs always well before the disk is fully resolved. In Be stars, the
  presence of a {small angular size} star contributing a significant amount of
  flux moves the threshold for complex phase behavior further toward a
  visibility squared of unity, {as seen in} Fig.~\ref{fig:interf}.

Based on our simulations, the astrometric regime is approximately valid up to
squared visibilities of $\approx0.8$, which corresponds to $\nu_{\rm obs}
\lesssim 1.5\rm\,m\,pc^{-1}$ for a typical Be star {(we note the correspondence of $\textbf{u}\cdot\textbf{r}$ with $\nu_{\rm obs}$; the higher $\nu_{\rm obs}$ the higher $\textbf{u}$).} The exact value of these
quantities depends on the target brightness distribution {and
  relative flux contributions}.  One example of the complex behavior outside
  the astrometric regime is reported by \citet{kra11}, who modeled AMBER/VLTI
  data for the Be star $\beta$~CMi. In this case, a phase reversal was
  registered at the DP signal of the longest baseline available for this
  nearby target (squared visibility of $\approx0.7$, $\nu_{\rm
    obs}\approx2.2$).

Outside the astrometric regime a phase reversal is an example of the phenomenology that emerges entirely from long-baseline interferometry, where the phase signal no longer corresponds to displacements of the photocenter. In the text that follows all interferometric computations are provided in the marginally resolved regime (i.e., when Eq.~\ref{eq:photc} is valid) to adequately separate the CQE-PS phenomenon from the just described effects. However, the CQE-PS signal can be identified and isolated even outside the astrometric regime. The validity and characteristics of the CQE-PS under these conditions are discussed in section~\ref{resolvedcase}.

\section{The CQE phase signature} \label{prfeat}

For the reference model described in Sect.~\ref{canon}, line emission from the rotating disk is the most important factor controlling the detailed shape of the DP. However, in some circumstances line absorption of photospheric light by the disk can also have a strong impact on these observables.

The spectroscopic phenomenon of the CQE was first predicted by
  \citet{han95} for the so-called shell-absorption lines in nearly edge-on Be
  shell stars. The CQEs consist of narrow absorption lines well below the
  stellar continuum that show a central weak cusp, known as CQE
  peaks. \citeauthor{han95} pointed out that these peculiar profiles result
  from the minimum at zero radial velocity in the absorption of the
  photospheric flux by the rotating CS gas. 
  However, \citeauthor{han95} expected them mainly in \ion{Fe}{ II } lines, which turned out to be
  formed at too large distances from the central star{,
where turbulence vs$.$ orbital motion becomes strong enough to disturb
the CQE signature}.  Later, \citet{riv99}
  identified these predicted CQE with previously unexplained
  observational features often seen in \ion{He}{ I } lines, and sometimes
  metallic shell lines.

\begin{figure*}[!ht]
\centering
\includegraphics[width=.31\linewidth]{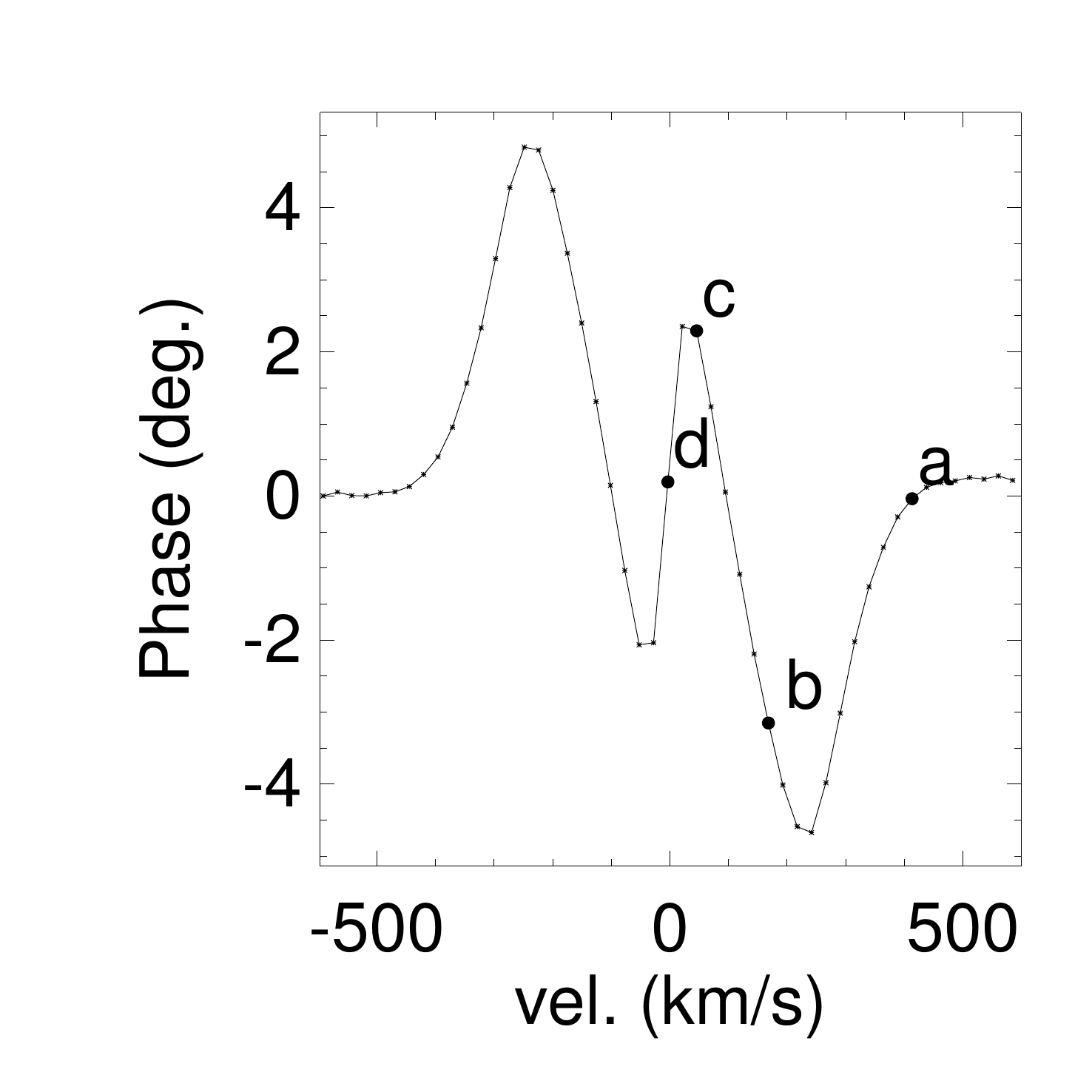}
\includegraphics[width=.62\linewidth]{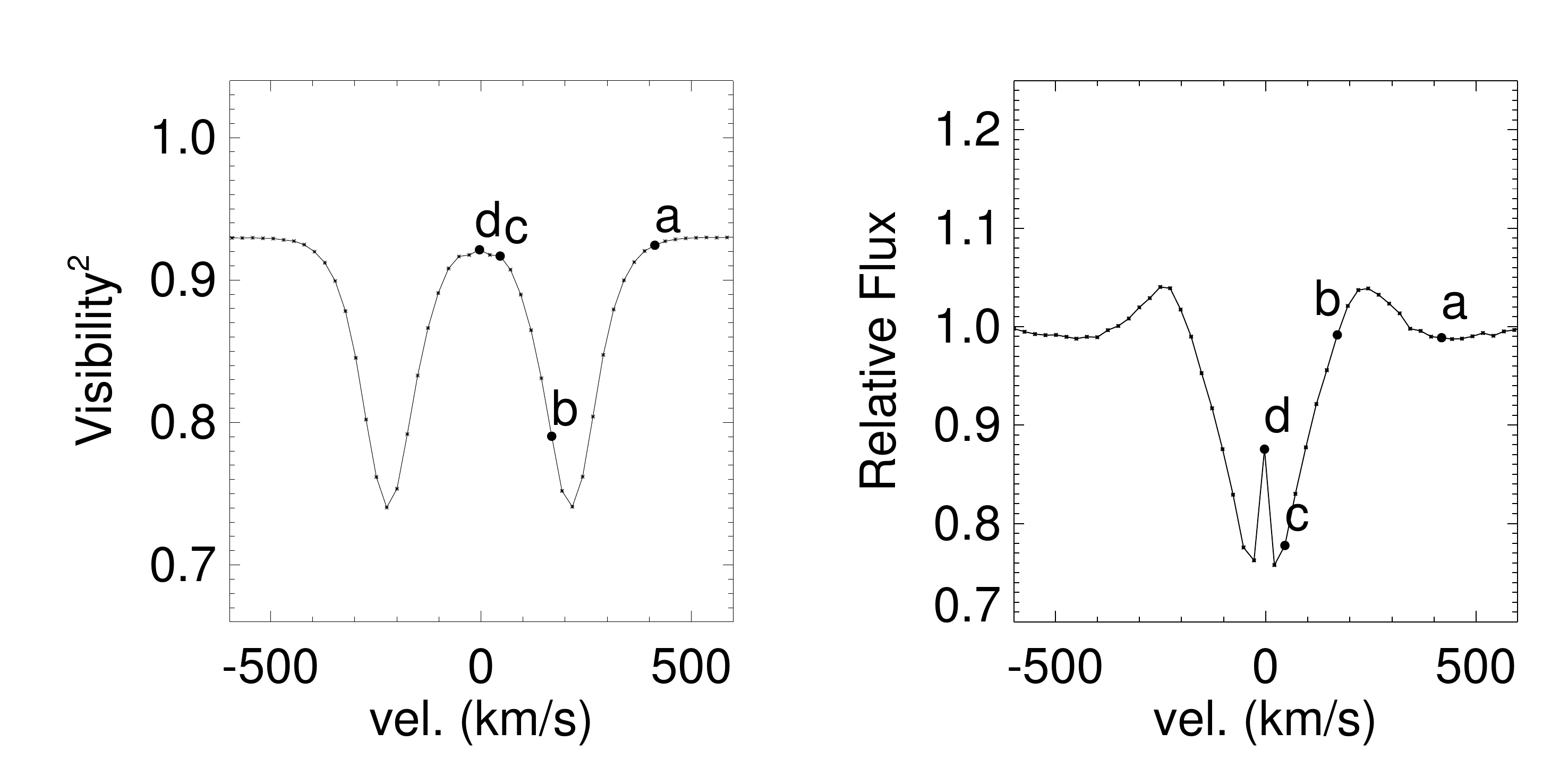} \\
\includegraphics[width=.9\linewidth]{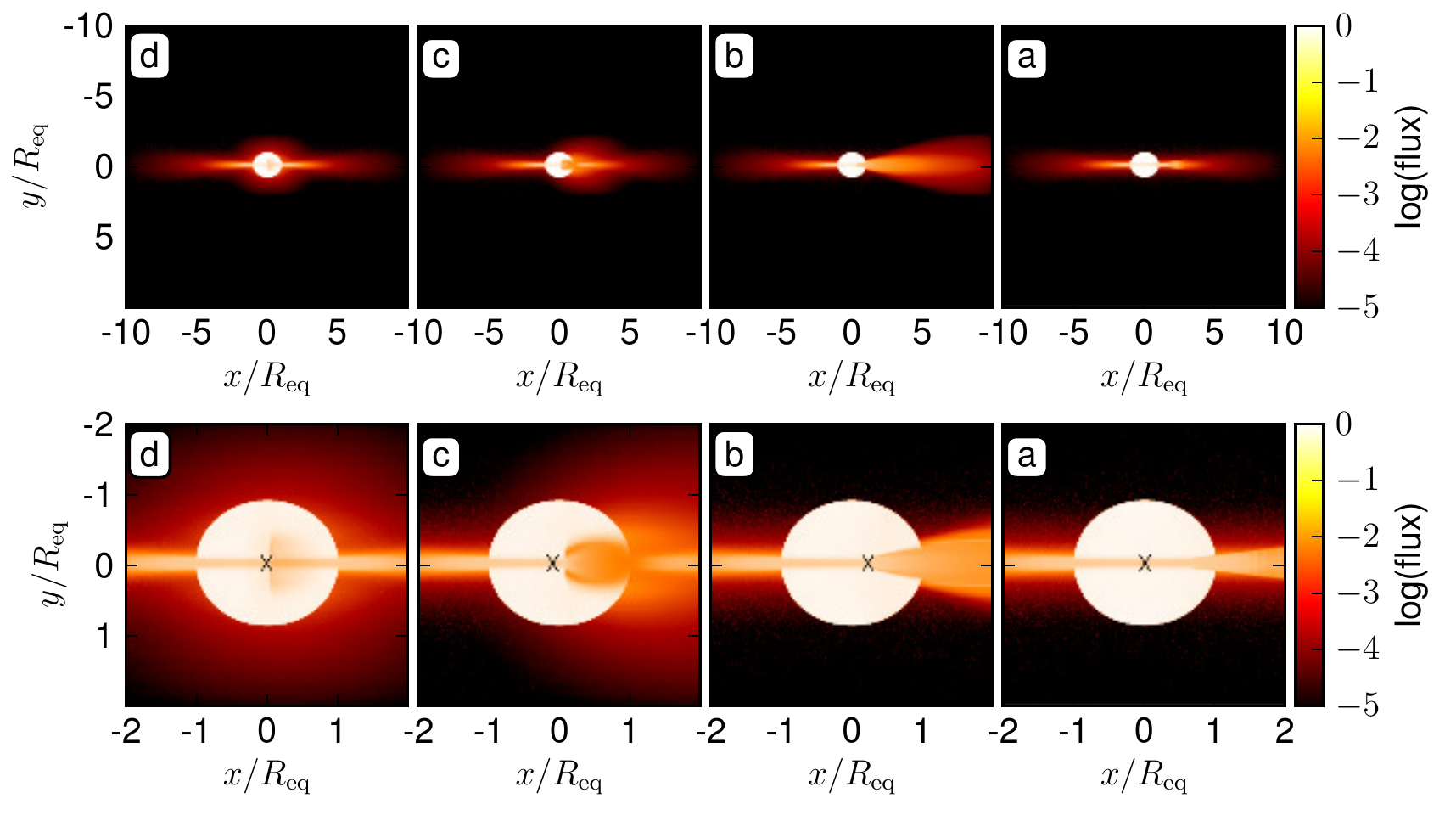} \\
\caption{Same as Fig~\ref{fig:classbe} for $i=90\deg$.}
 \label{fig:phot}
\end{figure*}

\begin{figure}
\centering
\includegraphics[width=.7\linewidth]{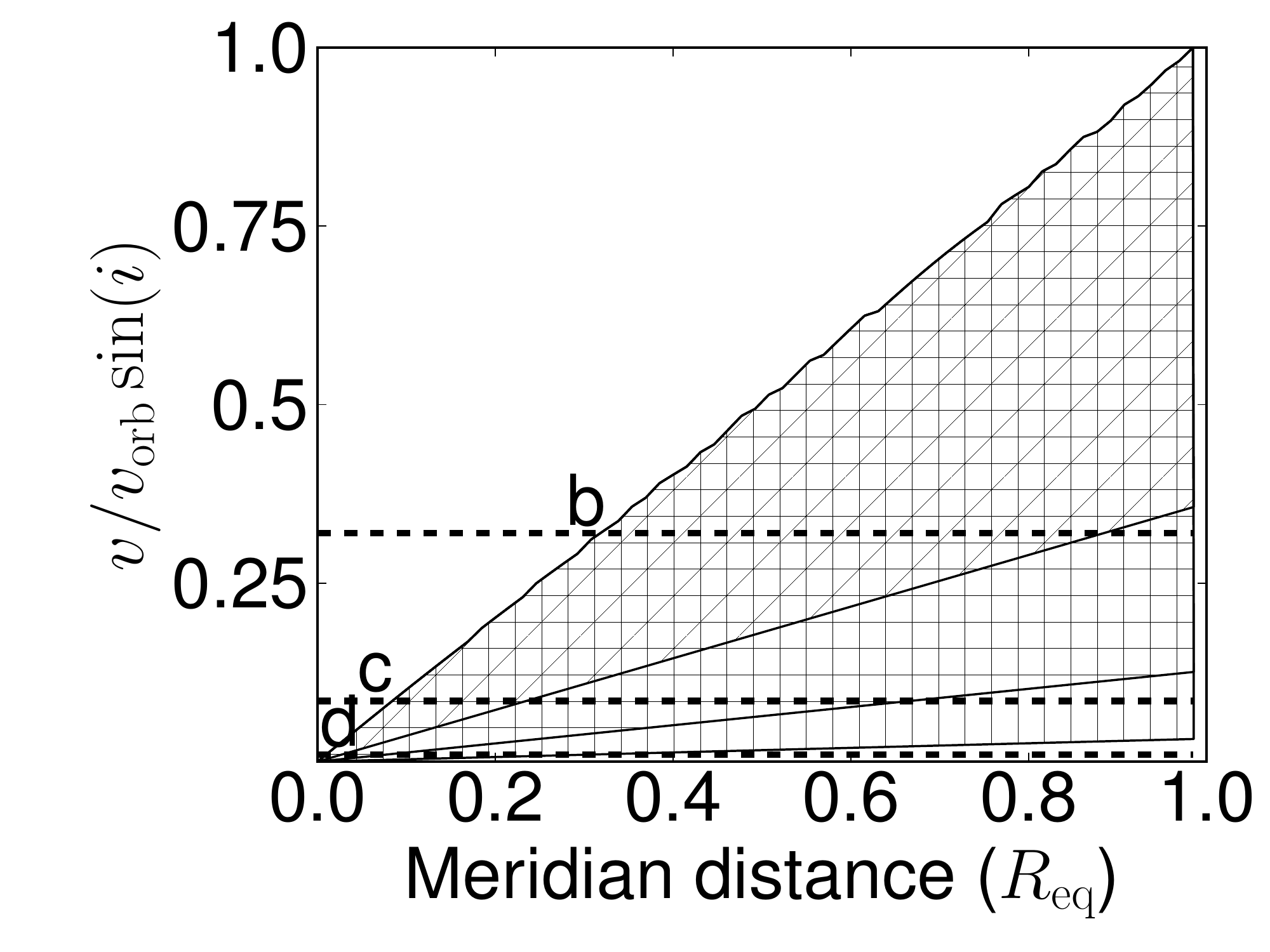} \\
\includegraphics[width=.7\linewidth]{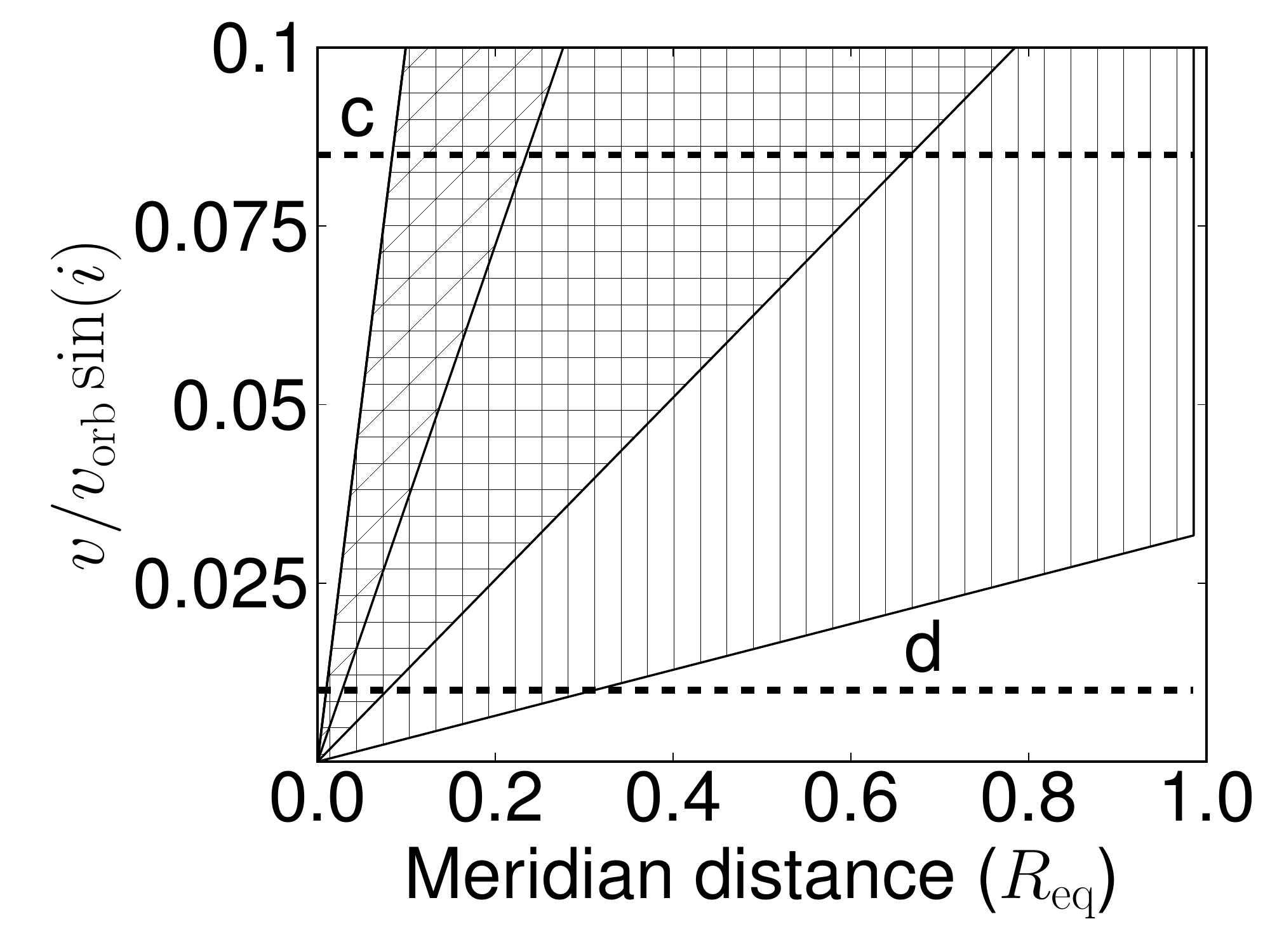} \\
\caption{{{
Projected velocity of a Keplerian disk along the line-of-sight in units of the projected orbital speed at the base of the disk ($v_{\rm orb}\sin i$) vs$.$ the normalized meridian distance (the distance to the stellar rotational axis in stellar radii). The intersection of a horizontal line with the hatched area indicates the range of meridian distances for which the projected velocity occurs.
Results for different disk sizes are plotted: $R_{\rm disk}=10\,R_{\rm eq }$, vertically hatched area; $4\,R_{\rm eq}$, horizontally hatched area; and, $2\,R_{\rm eq}$, diagonally hatched area. The velocities b to d from Figs.~\ref{fig:classbe} and \ref{fig:phot} are indicated as horizontal dashed lines.
\emph{Top}: full velocity range; \emph{bottom}: zoom of the low-velocity components. 
 } } }
 \label{fig:lambda_vs_p}
\end{figure}

This is better understood with the aid of
  Figs.~\ref{fig:phot} and \ref{fig:lambda_vs_p}, where we plot
  results for the reference case seen at an inclination angle
  $i=90^\circ$ (edge-on). At continuum wavelengths, the dark lane across the
  star is caused by continuum ({free-free} and bound-free) absorption in the
  CS disk (Fig.~\ref{fig:phot}, {velocity a}). For the short-wavelength range
  corresponding to the emission line profile, the continuum opacity may be 
  regarded as constant, so we do not expect to see changes in continuum
  absorption in the narrow-frequency range covered by Fig.~\ref{fig:phot}.  In
  spectral channels across the Br$\gamma$ line, additional line (bound-bound)
  absorption of photospheric light by \ion{H}{ I } atoms is seen
  (Figs.~\ref{fig:phot}, {velocities b to d}). This line absorption has a differential aspect
  because it depends on the line-of-sight velocity of the absorbing material
  {(see below)}. For high line-of-sight velocities, 
  $v_{\rm proj}$,
  only a minor part of the star is affected by {line} absorption
  (Fig.~\ref{fig:phot}, {velocity b}). Going towards lower velocities, i.e.,\ toward the
  line center, this fraction increases, until it reaches a maximum at about
  50\,km\,s$^{-1}$ (Fig.~\ref{fig:phot}, {velocity c}). The precise {value} depends on the
  outer disk size and density. 
  {{Closer to the line center}}, the {lowest} line-of-sight velocities are no
  longer absorbed
  (Fig.~\ref{fig:phot}, {velocity d}), giving rise to the spectroscopic CQE effect.  
Spectroscopic CQEs are thus, in spite of their name, not related to any emission process, but are an absorption phenomenon. 

\begin{figure}
\centering
\includegraphics[width=\linewidth]{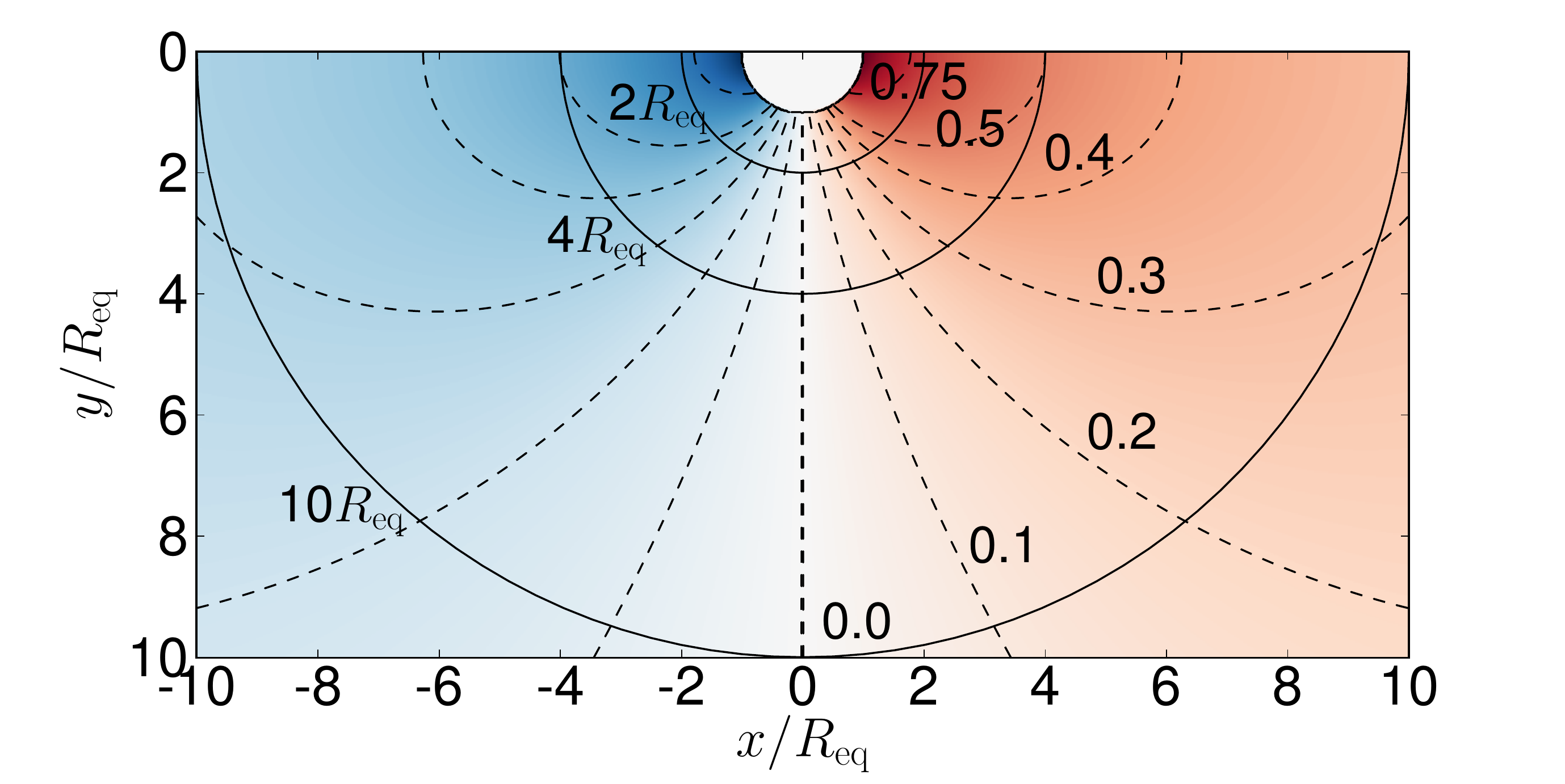} \\
\caption{{{
Isovelocity contours (dashed lines) for a Keplerian disk. Each curve corresponds to a fraction of the projected orbital speed at the base of the disk ($v_{\rm orb}\sin i$), as indicated.
The solid lines indicate different disk outer radii. 
 } } }
 \label{fig:isovel}
\end{figure}

{{Figure~\ref{fig:phot} can be understood with the aid of Fig.~\ref{fig:lambda_vs_p}.
We plot the 
projected velocity ($v_{\rm proj}$) in the line-of-sight  vs. distance from meridian. }
{{$v_{\rm proj}$ is expressed in terms of $v_{\rm orb}\sin{i}$, the projected Keplerian orbital speed at the base of the disk ($v_{\rm orb} = (GM/R_{\rm eq})^{1/2}$).}
For a given $v_{\rm proj}$, the figure shows the range of  distances from the meridian for which a given $v_{\rm proj}$ occurs along the line-of-sight}}. 
The fraction steeply increases from nearly zero at the meridian, where the line-of-sight velocity is zero, to a maximum close to $0.03\,v_{\rm orb}\sin(i)$ for the reference case (lower panel of Fig.~\ref{fig:lambda_vs_p}, vertically hatched lines). From this point on, the fraction decreases linearly with $v_{\rm proj}$, so that at high projected velocities only light that originated close to the edge of the star can be absorbed by the disk. 
The dearth of absorbers with zero velocity reduces the line depth, leading to the (false) impression of line emission. 
{{Fig.~\ref{fig:isovel} shows the isovelocity contours of a Keplerian disk.}
{The figure explains why}} the portion of the star covered by low-velocity particles is much larger for larger disks than for smaller ones.
For $R_{\rm disk} = 10\,R_{\rm eq}$, for instance, the largest surface coverage occurs for $v_{\rm proj} = 0.03\,v_{\rm orb}\sin(i)$ (lower panel of Fig.~\ref{fig:lambda_vs_p}). The projected velocity for which this occurs increases with decreasing disk radius as $(R_{\rm eq}/R_{\rm disk})^{3/2} v_{\rm orb}\sin(i)$. So, for $R_{\rm disk} = 2\,R_{\rm eq}$ the largest surface coverage will occur for $v_{\rm proj} = 0.34\,v_{\rm orb}\sin(i)$. This shows the importance of the disk size for the spectroscopic CQE.

This {line} absorption across parts of the stellar disk
generates a considerable decrease in the stellar flux, which in turn will
affect the photocenter position of the system, with a corresponding signal in
the DP. This is clearly seen in Fig.~\ref{fig:phot} as the ``wiggle'' in the center of the 
DP profile; the highest distortion is seen where the
  CS absorption is strongest, i.e., at about $50\,$km$\,$s$^{-1}$.
We observe the competition between {spectral line emission from
  the} disk at line-of-sight distances from the meridian higher
  than $1\,R_{\rm eq}$, which tends to shift the {monochromatic} photocenter towards one
side, and disk {spectral} line absorption {at line-of-sight distance from the meridian lower than $1\,R_{\rm eq}$}, which
shifts {the photocenter} toward the opposite side. So, at wavelengths
close to the line emission peak (Fig.~\ref{fig:phot}, {velocity b}), the disk emission
dominates over disk absorption and the photocenter position is shifted toward the emission lobe. For wavelengths closer to the line center
(Fig.~\ref{fig:phot}, {velocity c}), however, even though there is some disk emission, the
absorption is more important and the photocenter position is shifted to the
{opposite direction of disk emission}, with a corresponding change of sign of the DP. We conclude that the
\emph{spectroscopic phenomenon of the CQE has an interferometrical
  counterpart (dubbed CQE-PS}) that
presents itself as a central reversal in the DP profile. 
We note that this central reversal is an intrinsic phenomenon and is independent of the
reversals described in Sect.~\ref{nr_interf}. 

{We expect
that the interferometric CQE is more ubiquitous than its plain spectral counterpart.  Fig.~\ref{fig:lambda_vs_p} indicates that any radial motion above about
$0.03\,v_{\rm orb}\sin(i)$ would destroy the spectroscopic CQE signature {in a $10\,R_{\rm eq}$ disk}. This is true
regardless of the nature of this velocity, i.e., whether it is bulk
motion (e.g., out or in-flow) or generated by turbulence. We emphasize,
however, that the interferometric CQE signature is not
affected, this effect relies on obscuration that travels from one side of the star 
($-1>v_{\rm orb}\sin(i)>-0.03$) to the other ($0.03>v_{\rm orb}\sin(i)>1$),
while the spectroscopic CQE requires the lowest obscuration to occur at $v_{\rm proj}=0$.}

The effects of {photocenter displacement} can also be traced by the interferometric closure phases. In this work we only analyze the CQE effects on DP because they strongly rely on high spectral resolution data{, providing a direct measure of the target's photocenter over the projected baseline}. Accordingly, all discussions here are based on the photocenter of each spectral channel which is a quantity obtained by both techniques (closure phases and DP) in an equivalent way.

After understanding how disk absorption affects the DP, we now investigate under which circumstances the CQE-PS should be observable  and what its diagnostic potential is for constraining the physical properties of the Be + disk system.

\subsection{Spectral resolution \label{sec:specres}}
The CQE-PS critically depends on the spectral resolution of the interferometer, because the differential absorption that causes the CQE is only strong in a narrow wavelength range.

\begin{figure}[!ht]
\centering
\includegraphics[width=.7\linewidth]{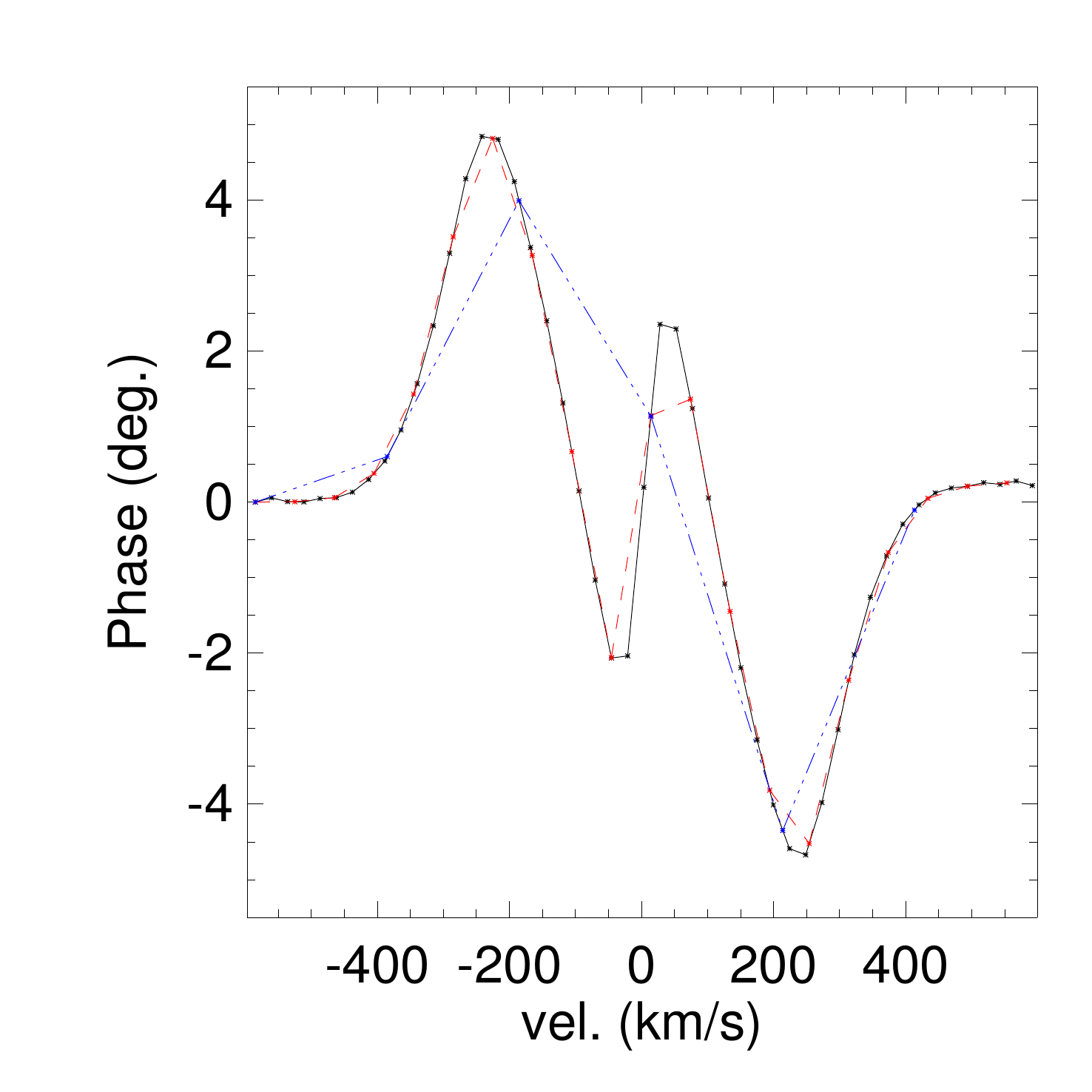}
\caption{Differential phase signal for the Be reference model with different spectral resolutions; $R=12000$ black (full), $R=5000$ red (dashed), and $R=1500$ blue (dot-dashed) lines. Disk inclination angle is $i=90^\circ$.}
 \label{fig:specres}
\end{figure}

In Fig.~\ref{fig:specres}, we see simulations of observations with different
spectral resolutions for our reference model seen at $i=90^\circ$. The figure
shows that the phase reversal would be unobservable at $R\approx1500$,
{which is} the medium resolution of the AMBER
interferometer.
Broader spectral channels also change the values of the DP
since different emitting regions are sampled in each channel. 

\subsection{Geometrical considerations}

Even if the physical conditions are ideal for CQE-PS, its occurrence strongly depends on the angle $i$ and less intensely on the baseline on-sky orientation. 

\begin{figure}[!ht]
\centering
\includegraphics[width=.7\linewidth]{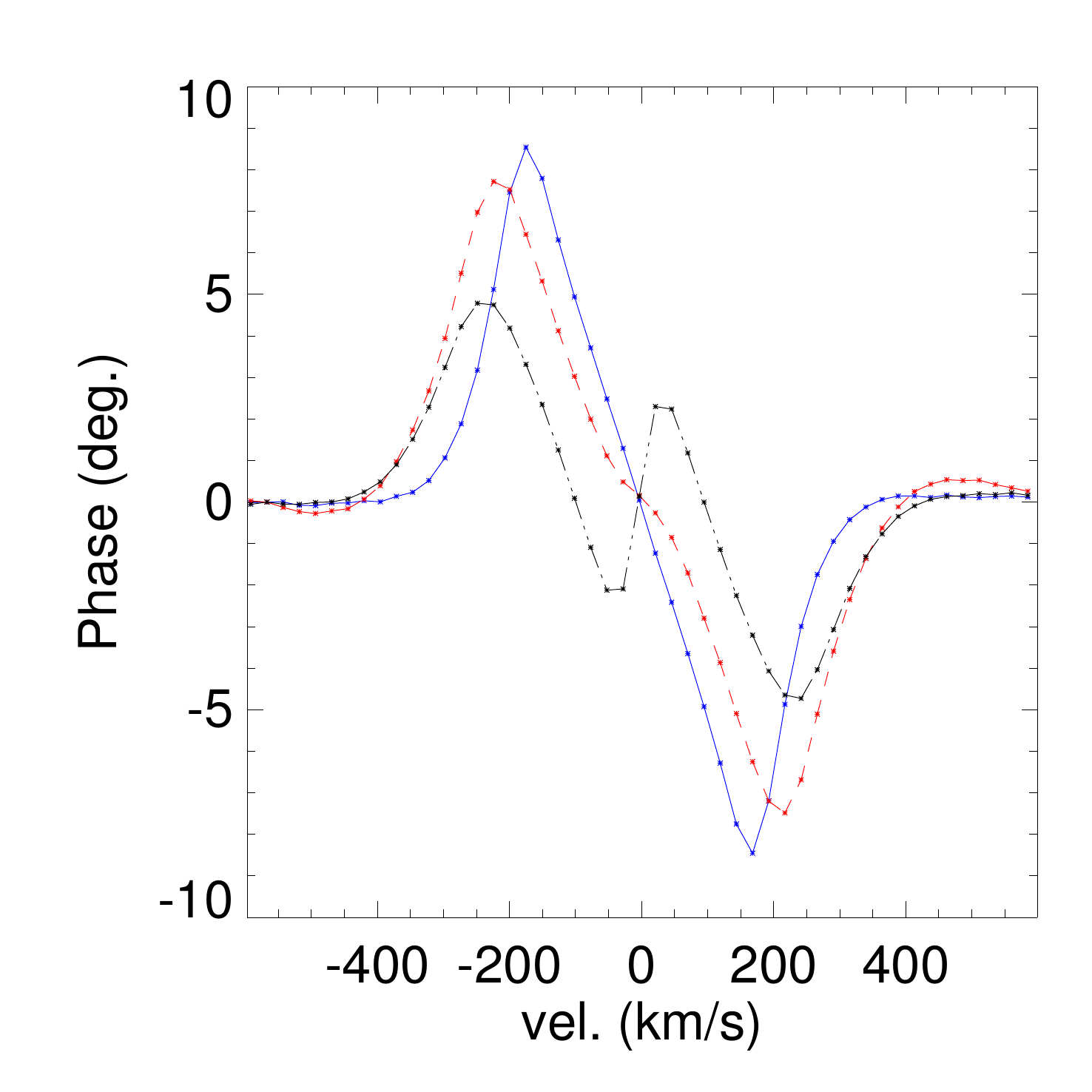}
\caption{Differential phase signal for the Be reference model 
{for different disk inclinations}: $i=90^\circ$ black (full), $i=75^\circ$ red (dashed), and $i=45^\circ$ blue (dot-dashed) lines.}
 \label{fig:f_anglei}
\end{figure}

As an absorption effect, the CQE-PS will be stronger when the system is viewed
edge-on, since the densest regions of the disks are those closest to the
stellar equatorial region. Clearly, when the disk is viewed face-on, this
phenomenon will not {occur, since there is no absorbing material in front of
  the star.} The effects of $i$ on the CQE-PS are shown
in Fig.~\ref{fig:f_anglei}, which demonstrates that the photocenter
displacement due to the differential absorption is gradually weaker because less
photospheric flux is absorbed for lower $i$.

\begin{figure}[!ht]
\centering
\includegraphics[width=.7\linewidth]{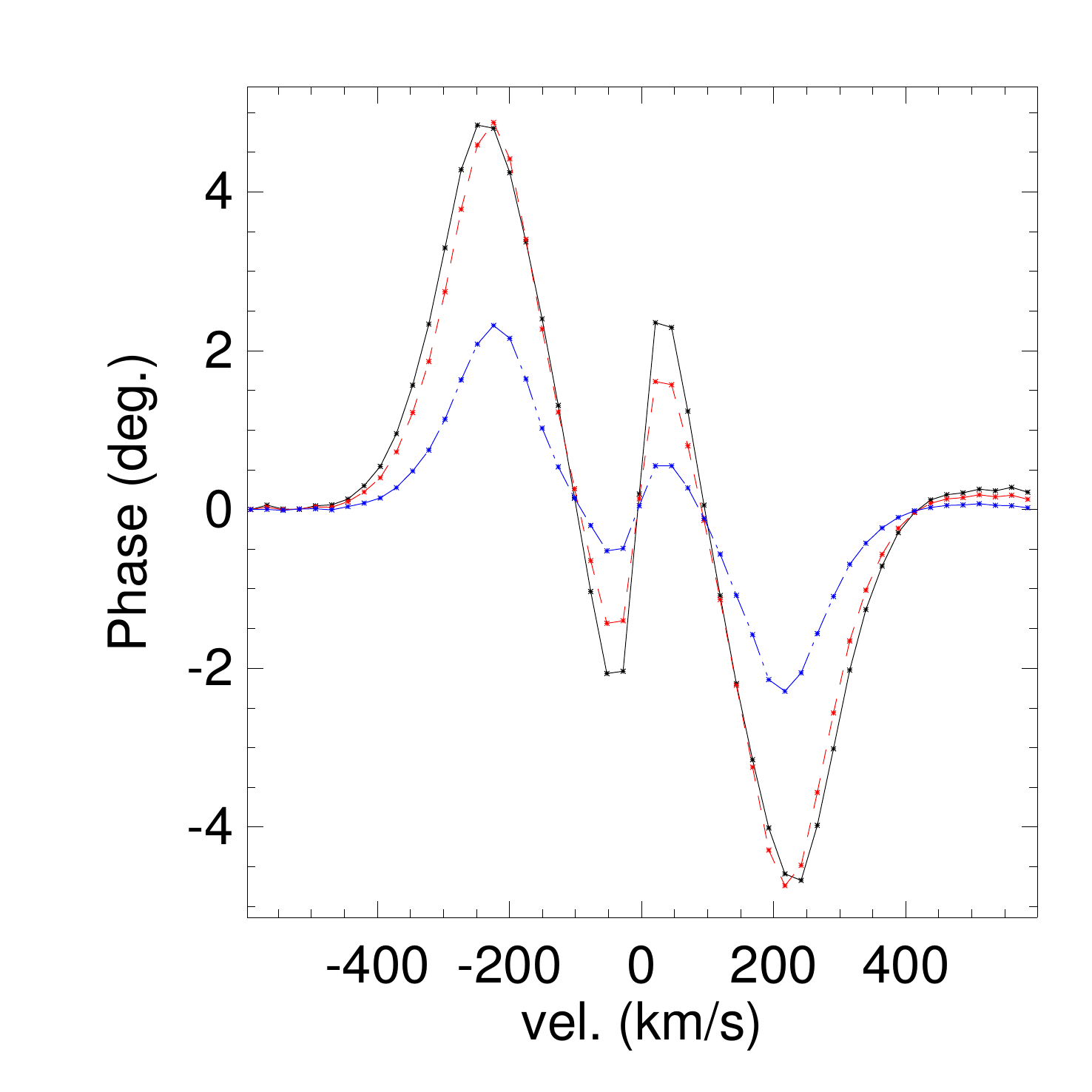}
\caption{Differential phase signal for the Be reference model for {different angles between the projected baseline and the on-sky disk orientation: $\theta=0^\circ$ black (solid), $\theta=45^\circ$ red (dashed), and $\theta=75^\circ$ blue (dot-dashed) lines. Disk inclination angle is $i=90^\circ$.}}
 \label{fig:mod3b_pa}
\end{figure}

Another factor that contributes to the amplitude of the CQE-PS is the baseline position angle. Because the photocenter displacements are expected to be parallel to the disk orientation, the CQE-PS will be highest when the baseline orientation is parallel to the disk major axis, as illustrated in Fig.~\ref{fig:mod3b_pa}.

\subsection{CQE-PS outside the astrometric regime} \label{resolvedcase}

\begin{figure}[!t]
\centering
\includegraphics[width=.7\linewidth]{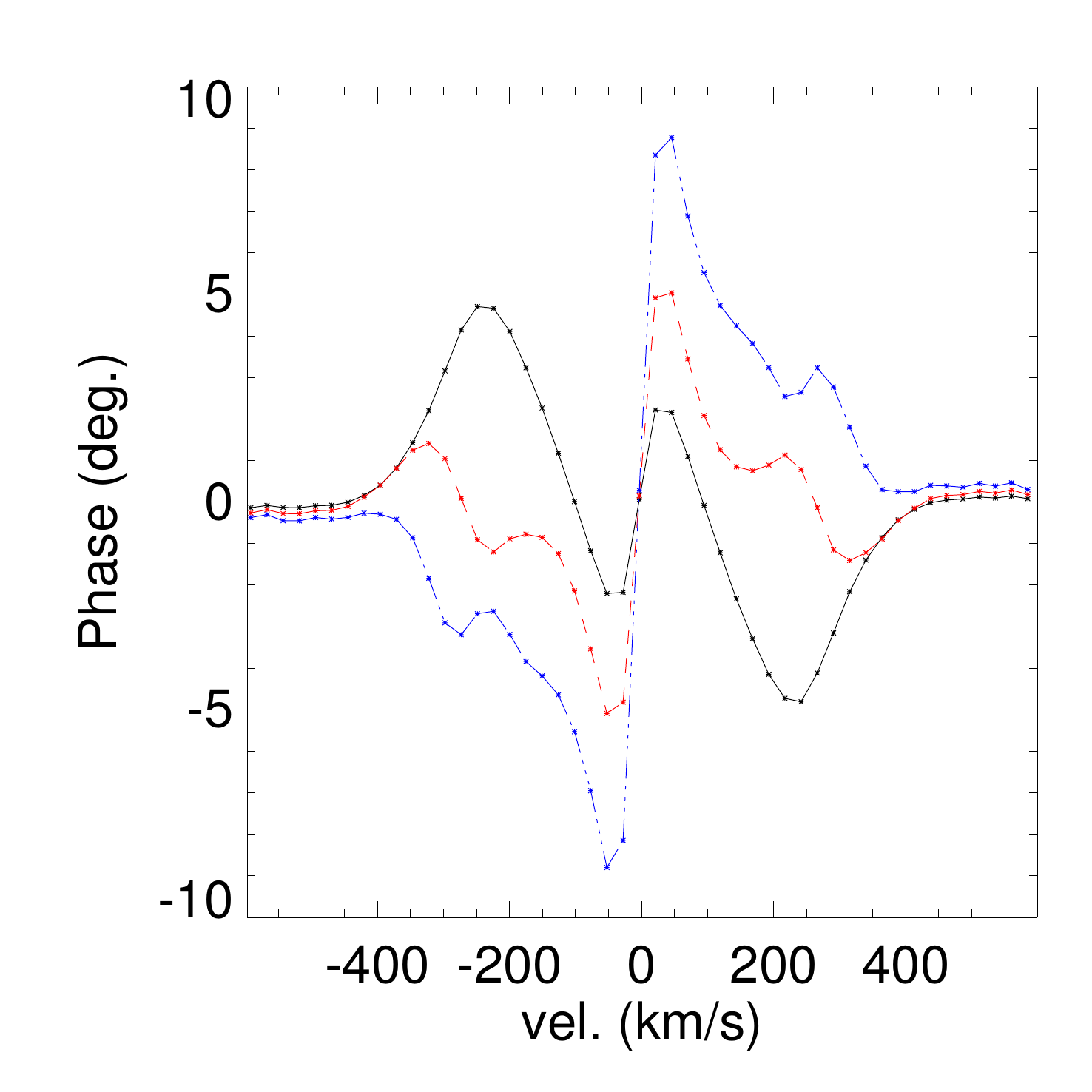}\\
\includegraphics[width=.5\linewidth]{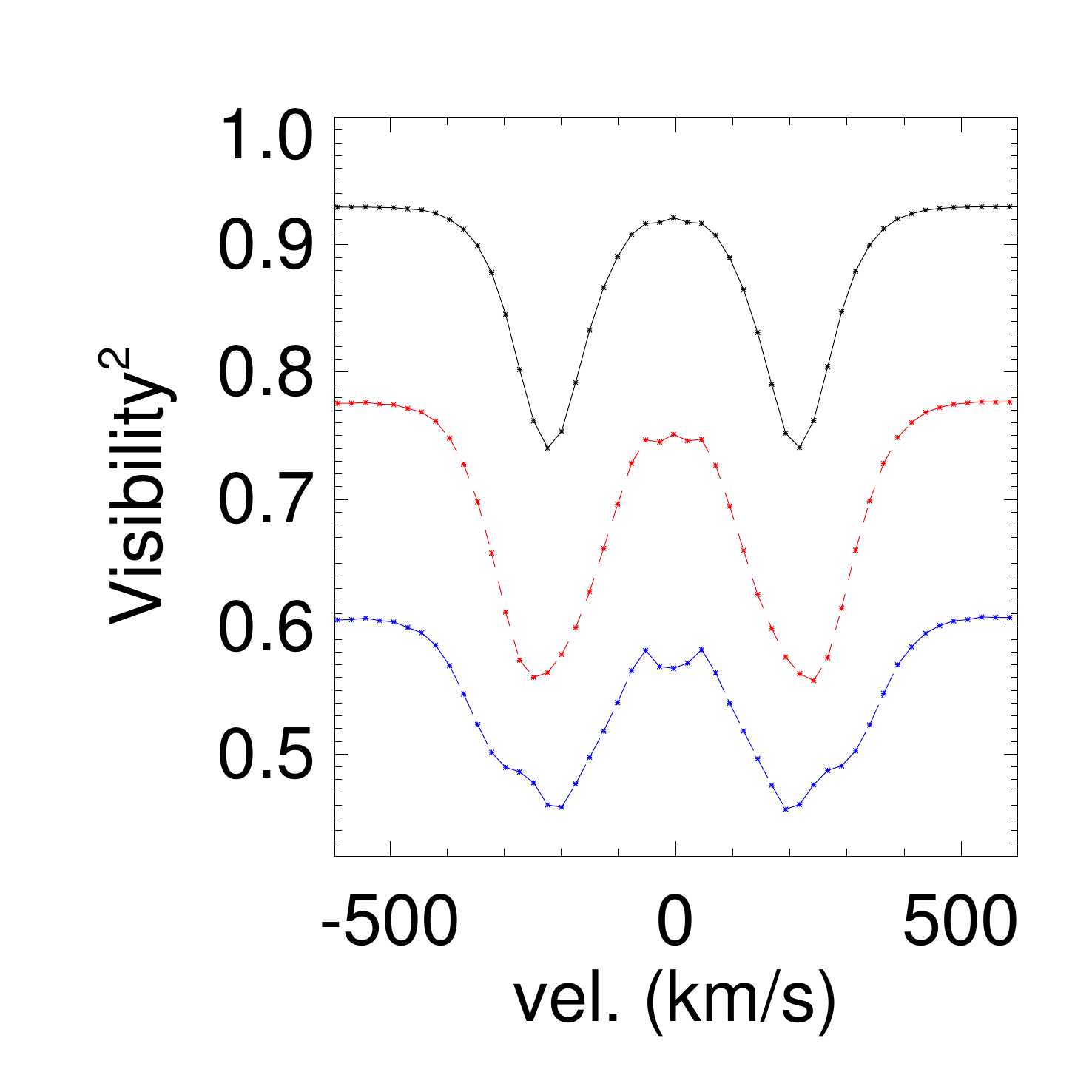} \\
\caption{\emph{Top}: The differential phase signals for the Be reference model at $i=90^\circ$. The reference model is the black (full) line, observed with $\nu_{\rm obs}= 1\,\rm m\,pc^{-1}$ (astrometric regime); red (dashed) line $\nu_{\rm obs}=2\,\rm m\,pc^{-1}$ and blue (dot-dashed) line $\nu_{\rm obs}=3\,\rm m\,pc^{-1}$ (complex phase behavior).
\emph{Bottom}: The corresponding visibility signal.}
 \label{fig:unres2}
\end{figure}

The CQE-PS so far has been analyzed in the astrometric regime {for which there is a direct correspondence between the target's photocenter and the DP. However, the CQE-PS is also present outside this regime, and the combination of the CQE-PS with high-resolution phase effects can result in very complex phase profiles. For instance, Fig.~\ref{fig:unres2} shows how the DP signal is affected by higher spatial resolution: while the high-speed components display a complex behavior due to its more extended associated on-sky size, the CQE-PS part of the profile (low-velocity components) is still in the linear regime. Therefore, at high angular resolution the CQE-PS may even come to dominate the phase signal, which documents a phase reversal}. 

\section{Diagnostic potential} \label{param}
In this section we investigate the effects of several physical parameters on the amplitude and shape of the CQE-PS. Hereafter, we assume that the interferometric baseline runs parallel to the stellar equator and $i=90^\circ$, since this choice maximizes the CQE-PS amplitude. {All results shown were calculated {for the reference model} for $\nu=1\,m\,pc^{-1}$, which corresponds to a typical Be star of spectral type B1\,V seen at a distance of 100 pc and observed with baselines of 100 m}.

In addition to the disk size, density slope, and density scale, we also explored the effects of stellar rotation rate and spectral type, disk scale height, and turbulent speed on the CQE-PS. We found that the shape of CQE-PS is only little affected by these last parameters.

\subsection{Disk density}

\begin{figure}[!ht]
\centering
\includegraphics[width=.7\linewidth]{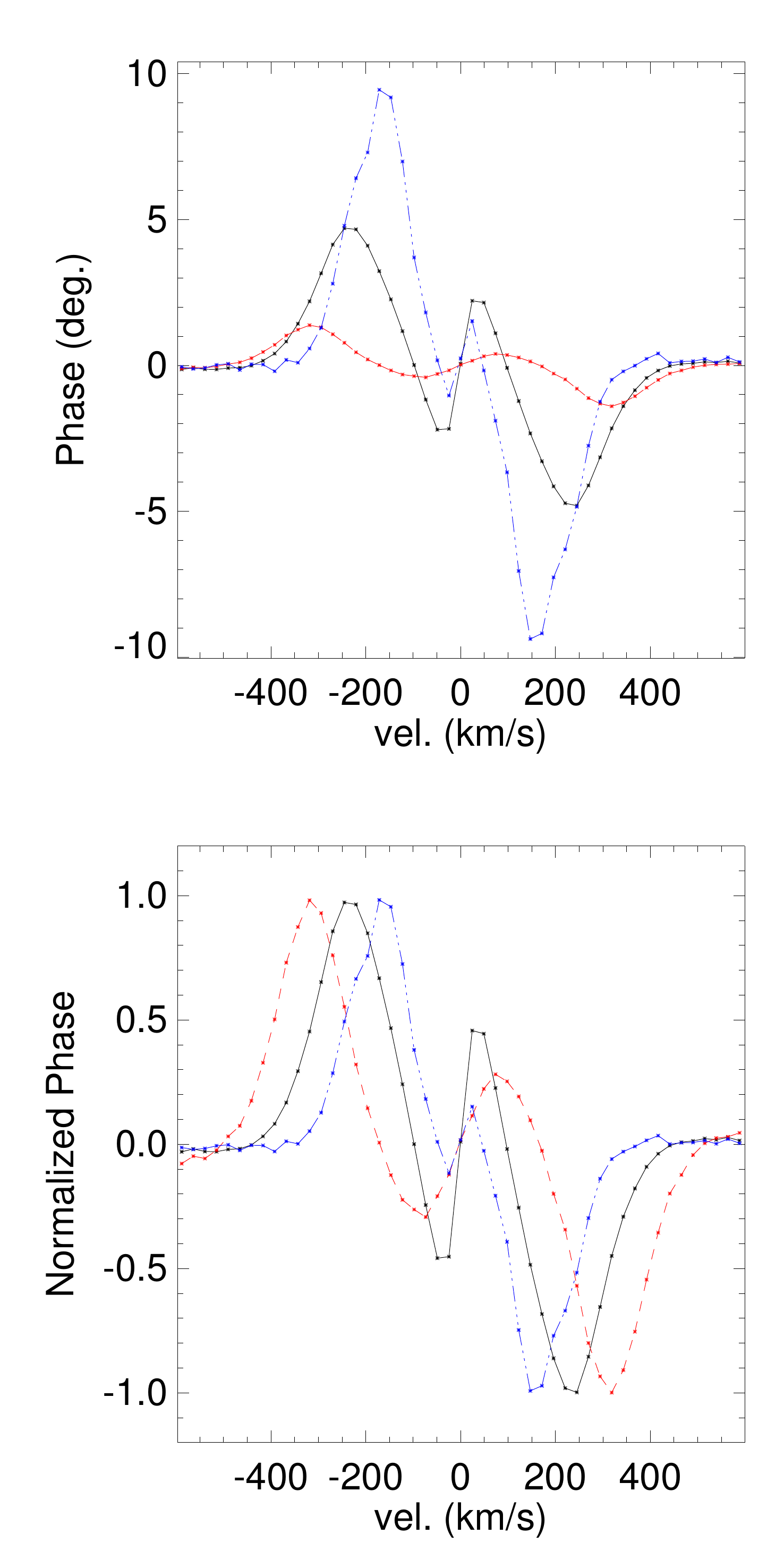}
\caption{Differential phase signal for the Be reference model for low-  ($10^{12}\,\rm cm^{-3}$, red dashed line) and high-density ($10^{14}\,\rm cm^{-3}$, blue dot-dashed line) models. The reference model is the intermediate value (black full line).}
 \label{fig:dens}
\end{figure}

{The} CQE-PS is a disk absorption effect,
observable in the DP of Be stars together with the emission signal from the
disk. The CS disk density is an important quantity for characterizing {the stellar mass-loss rate and drives} both the absorption and emission processes by the disk.

One can conceive two limits to the CQE-PS occurrence in terms of the disk density. Assuming a tenuous disk, absorption (both continuum and line) probably does not block the stellar photospheric flux very much, and we expect the phase profile to be little affected by it. 
{At the other extreme,the opacity becomes so intense for very large disk densities that absorption is high, even at continuum wavelengths. However, emission is also quite high at high densities, therefore the occurrence of the CQE-PS in the high-density regime will depend on the process (absorption vs$.$ emission) that controls the photocenter position at the wavelengths close to the line center}.

{In the top panel of Fig.~\ref{fig:dens} the effects of disk density on the CQE-PS are shown in a quantitative way. We considered three values for the number density at the base of the disk $n_0$: $10^{12}\,\rm cm^{-3}$, $10^{13}\,\rm cm^{-3}$ (the reference model), and $10^{14}\,\rm cm^{-3}$. Since the DP amplitudes of these models are very different, the DPs were rescaled in the bottom panel to better illustrate the relative effect of the differential absorption versus differential emission on the shape of the DP. 
The relative weight of absorption vs$.$ emission for different densities causes considerable changes in the detailed shape of the DP, which indicates that the CQE-PS can be a good quantitative diagnostic of disk density. For instance, in the high-density model (blue curve in Fig.~\ref{fig:dens}) the emission almost completely outweighs the absorption, and as a result, the DP profile has just a slight hint of the CQE-PS.
In our modeling we found that for a typical Be star, the density value that results in the highest contrast between the emission and absorption (i.e., the most pronounced CQE-PS) is $n_0\sim10^{13}\,\rm cm^{-3}$.}

\subsection{Disk size}  
{In this subsection, we evaluate the effect of the disk outer radius on the CQE-PS. Very little is known about the actual size of Be star disks. For known Be binaries, such as $\zeta$~Tauri \citep{ste09}, the disk is expected to be truncated at some radius \citep[see, e.g.,][]{car09}, but for isolated stars there is in principle no limit for the disk size \citep{oka01}.}

\begin{figure}[!ht]
\centering
\includegraphics[width=.7\linewidth]{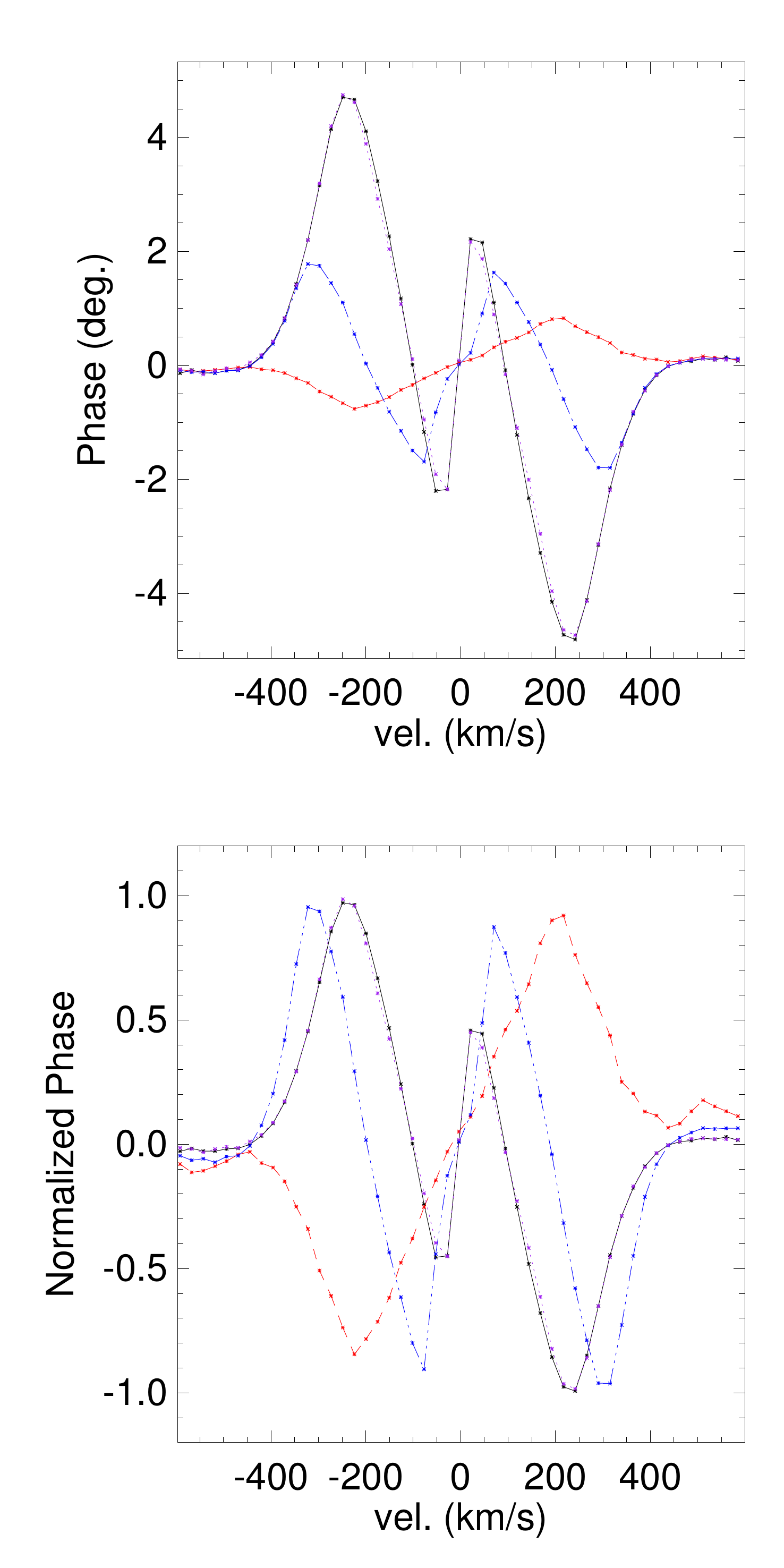}
\caption{Differential phase signal for the Be reference model for different
  disk sizes: $R_{\rm disk}=2$ (red dashed line), 4 (blue dot-dashed line), 10 (black full line) and $20\,R_{\star}$
  (purple pointed line). {The signals for $R_{\rm disk}=10$ and $20\,R_{\star}$ are almost coincident. }
}
 \label{fig:rdisk}
\end{figure}

{Alternatively, one can define a region of the disk that is radially opaque to continuum or line radiation and therefore dominates the disk emission. The radius of this \emph{pseudo-photosphere}}\footnote{Note that other works (e.g., \citealp{har94, har00}) have used the term pseudo-photosphere to denote the region within which the disk viewed face-on is optically thick and so looks like an extension of the stellar phtosphere.}{($R_{\rm p}$) depends on the wavelength considered;} 
for instance, as discussed in \citet{hau12}, for a fully developed disk, the pseudo-photosphere extends to about $R_{\rm p}=2\, R_{\star}$ at the $V$ band and $R_{\rm p}=4\, R_{\star}$ at the $K$ band. For
line transitions such as \ion{H}{ I } Br$\gamma$, this region can be as large as  $R_{\rm p}=10$--$20\, R_{\star}$. 
{If the physical size of the disk is larger than the radius of the pseudo-photosphere at a given band, observations in this band constrain only the {latter, and not the former.}
This point is illustrated in Fig.~\ref{fig:rdisk}, where we present simulations of the reference model with different truncation radii ($R_\textrm{disk}=2$, 4, 10 and $20\,R_{\star}$). The models with the largest disk radii are essentially identical, which means that the size of the pseudo-photosphere for the reference model is $R_{\rm p}\lesssim 10\,R_{\star}$. }

{The situation for small disk models is quite the opposite; as shown in Fig.~\ref{fig:rdisk}, the CQE-PS provides an unequivocal constraint to the size of the disk. {Insofar} as we consider smaller disks, the size of the emitting lobes decreases; as a result, the central reversal of the DP profile, which is a measure of disk absorption, becomes increasingly more marked (see the red curve in Fig.~\ref{fig:rdisk}, which shows results for  $R_\textrm{disk}=4\,R_{\star}$). The net effect is that the DP peak shifts away from the line center, because the low-velocity components of the disk were removed, and the amplitude of the central reversal increases. Interestingly, for very small disks the central reversal may even dominate the entire phase profile, providing what looks like an inverted phase profile. Observations of such a disk might be misinterpreted as a disk rotating in the opposite direction! Note, however, that disk emission for such a small disk still appears as small ``wings'' in the DP profile where no significant absorption is present and the disk emission can be clearly seen.}

\subsection{Radial density distribution} 

\citet{hau12} studied the temporal evolution of viscous disks around Be stars
and how different {mass-loss rate histories} affect the disk structure. It was
found that the disk density profile is highly variable in time and that is is
not possible to assign one single density slope $m$ ($n\propto r^{-m}$) to
the entire disk. However, constraining the disk slope from observations can be
quite useful, since an actively forming disk will have quite different density
slopes than dissipating disks. For the former case, $m$ is typically larger
than 3.5, while for the latter it is $m \lesssim 3$. 

\begin{figure}[!ht]
\centering
\includegraphics[width=.7\linewidth]{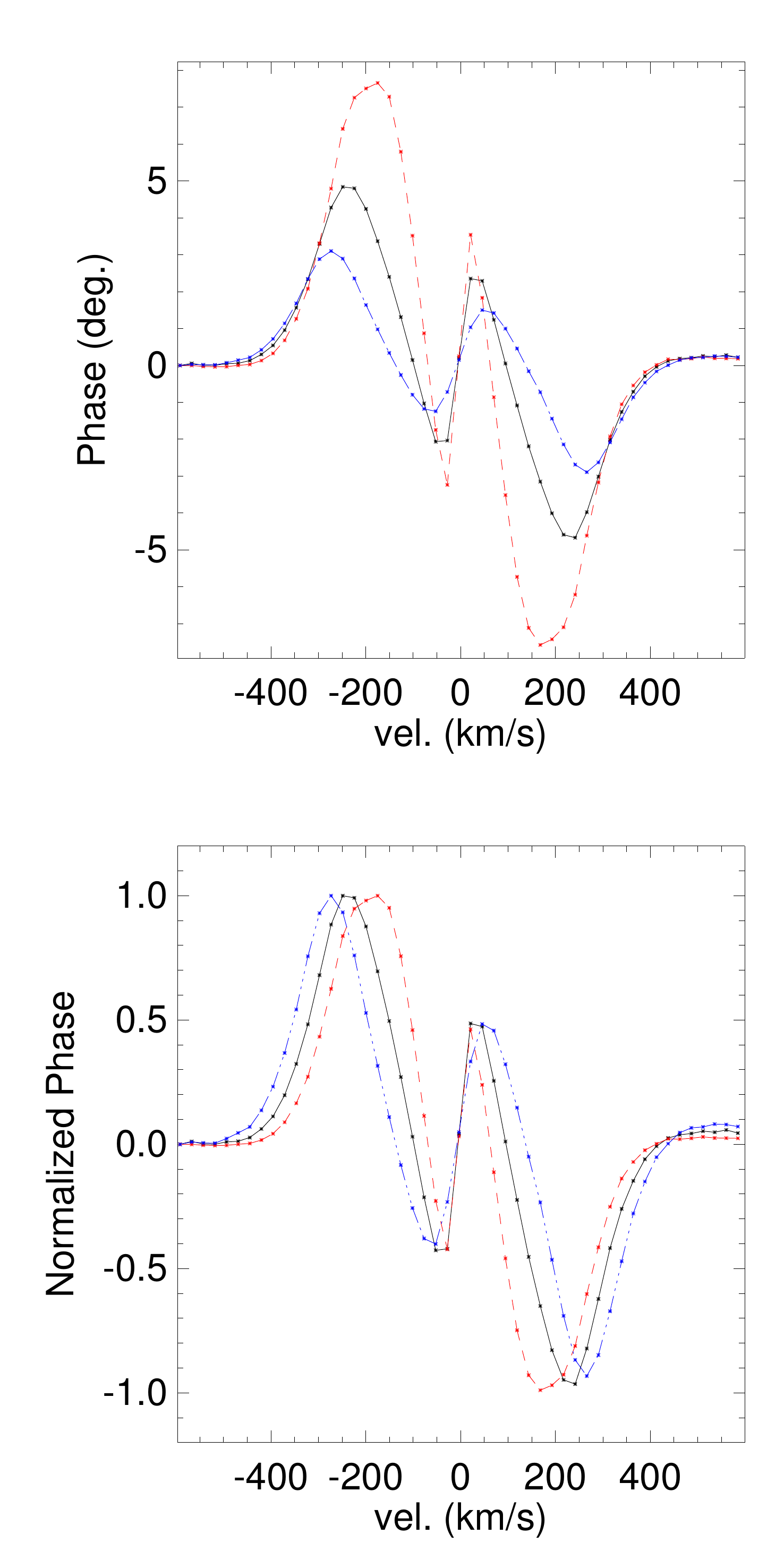}
\caption{Differential phase signal for the Be reference model with different radial density exponents: $m=3.0$ {(red dashed line)}, 3.5 (black full line) and 4.0 {(blue dot-dashed line)}. 
}
 \label{fig:dens_n}
\end{figure}

Here again the pseudo-photosphere concept can be useful. Flat disk-density
distributions ($m < 3.5$) correspond to larger pseudo-photospheric radii,
which lead to higher contributions of lower projected velocities in the
phase profile, while the converse is true for sharp disk-density profiles
($m>3.5$).  Fig.~\ref{fig:dens_n} shows the Be reference model run for
different radial density exponents.  As the disk density slope decreases, the
density becomes concentrated in the inner regions and the CQE-PS reversal
feature broadens and increases in importance relative to the emission
part of the phase profile.

{Both a flat disk density distribution (low index $m$) and a high base density ($n_0$) enhance the disk emission in a similar way, thus creating large PS amplitudes (Figs.~\ref{fig:dens} and \ref{fig:dens_n}). However, the CQE-PS may allow distinguishing the two cases by comparing the detailed shape of the central reversal: whereas a higher density decreases the amplitude of the phase reversal, a lower index $m$ increases it. This arises from the fact that for lower-density exponents the opacity decreases more slowly with radius. Therefore, a flat density slope is associated with more absorbing material in front of the star, which results in stronger  contrasts between the flux from one side of the star vs$.$ the other. This is illustrated in Fig.~\ref{fig:lambda_vs_p}, which shows that the larger the size of the disk (or, equivalently, of the pseudo-photosphere), the larger the amount of material with the appropriate projected radial velocity.}

\subsection{Asymmetric configurations and time-dependence}

{The CQE-PS can be also employed to trace short-time CS variations thanks to its sensitivity to the disk inner regions. 
This information, linked to disk rotation, enables a disk azimuthal characterization. Similarly, it can map the evolution of the disk properties as a function of time.}

{The characterization of disk properties at different rotational phases would be useful for studying complex phenomena, such as the so-far unexplained triple-peaked H$\alpha$ emission profiles (e.g., $\zeta~\rm Tauri$, as in \citealt{ste09}). This particular feature may be restricted to binary Be systems, but it is not well understood. Determining the disk configuration in such a system could offer some constraints for the influence of the companion.
}

{Moreover, when Be stars change their brightness, this can originate in the star, the disk, or both. Although most of the observed variations are attributed to disk changes \citep{hau12}, the CQE-PS can be used to settle this question because it is very sensitive to the disk part whence most of the excess flux in the visible range comes from.}

\subsection{Stellar angular size}

Because CQE-PS absorption is generated across the stellar photosphere, it can be used to estimate the stellar angular size from the CQE-PS signal.

{In a rotating disk, the positive (or negative) radial velocity components are projected at half of the stellar photosphere, which means that differential absorption can block at most half of the photospheric flux at a given wavelength. It can be easily demonstrated in this scenario of maximum absorption that the displacement of the photocenter due to absorption is a fraction {$p\approx2/5$} of the stellar radius, {where $p$ is the barycenter of a unitary and uniform vertical semi-circle: }

\begin{equation}
p = \frac{2}{\pi}\int_0^1\int_{-\sqrt{1-x^2}}^{\sqrt{1-x^2}}x\,dy\,dx \,.
\label{eq:p}
\end{equation}

The photocenter shift (Eq.~\ref{eq:photc}) driven by absorption can be measured as  
$\Delta\boldsymbol\epsilon=p\cdot({R_{\rm eq}/d})$, where $R_{\rm eq}$ is the stellar radius, and $d$ its distance with $0<p<2/5$. This quantity is particularly interesting because we have an estimate of the ratio of $R_{\rm eq}/d$ in a visibility-independent measurement. 

In our reference model, which can be considered representative for the stars that exhibit the phase-reversal profile, the factor $p$ is about 1/8. This value, lower than $2/5$, is due to a smaller obscured area of the star and to the contribution of line emission by the disk. In any event, this analysis allows one to obtain a lower limit of the stellar radius. 

\section{Conclusions} \label{concl}

{
With the advent of modern spectro-interferometers it became possible to measure the variation of the interferometric phase across an emission line. It was generally expected that these phase diagrams would display a simple S-shaped profile consisting of a single reversal, resulting from line emission from a rotating disk. This picture was largely confirmed by measurements made with moderate spectral resolution.
However, at high spectral resolution, such as can be obtained by AMBER/VLTI, observations have shown a more complex phase profile. 
Conceptually, deviations from the canonical S-shaped profile can be expected in two circumstances:}
{(i) by purely interferometric properties of observations outside the marginally resolved regime, where the differential phases no longer correspond to the photocenter of the system (i.e., end of the astrometric regime), and (ii) by physical processes accessible at high spectral resolution that affect the interferometric signal. 
We found for a typical Be system that the astrometric regime occurs when the visibilities are higher than 0.8. This corresponds to observations when $\nu_{\rm obs}$ is lower than $1.5\rm\,m\,pc^{-1}$. The exact value of this
quantity depends on the target's brightness distribution.}

{We discussed the role of radiative transfer effects (ii, above), in particular, line transfer in a rotating disk, on determining the detailed shape of the high-resolution phase signal.
We demonstrated that differential line absorption by the rotating disk can cause a well-defined photocenter displacement, which in turn affects the detailed shape of the interferometric phases. We dubbed this effect CQE-PS (central quasi-emission phase signature), in reference to the fact that the same mechanism produces the CQE phase profile of shell stars \citep{han95,riv99}.
}
A detailed explanation of the CQE-PS was presented in Sect.~\ref{prfeat}.

{The shape and amplitude of the CQE-PS depends on a number of parameters, both intrinsic to the system and related to the observational setup (baseline length and orientation)}. 
{Using a realistic model consisting of a rotating B1\,V star with a viscous decretion disk, we made an initial study to assess to diagnostic potential of CQE-PS.}

{Among the several model parameters explored, we found that the signal is 
sensitive to the CS disk density, disk size, 
and the radial density distribution of the CS material. Varying each of these parameters generated a characteristic shape in the differential phase diagram, which demonstrates the diagnostic potential of the CQE-PS. The amplitude of the predicted signals in phase is higher than the error for the current interferometers with high spectral resolving power (typically $\lesssim 1^\circ$ for AMBER/VLTI and $\lesssim  3^\circ$ for VEGA/CHARA).}

{Certain conditions and parameters make the effect more pronounced. Observationally, the effect is strongest for edge-on viewing (shell stars) and for baselines oriented parallel to the disk equator. Physically, the effect is strongest for base disk densities of $n_{0}\sim10^{13}\,\rm cm^{-3}$ and disks larger than about $10\,R_{\star}$.}

{A remarkable result of the CQE-PS is its ability to make a (lower) estimate of the stellar angular size even when the observed squared visibilities are close to unity. This result may extend the distance range at which such an estimate can be made with interferometry.} 

{The full power of this diagnostic tool will be realized when it is applied to observations over a full disk life-cycle, from first ejection through final dispersal, over a full $V/R$ cycle of disk oscillation, or over a full orbital period in a binary system. The determination of the temporal dependence of the CS parameters will enable exploring fundamental physical processes operating in the disk, which are responsible for the disk evolution timescales, the stellar mass-loss rate, and mass ejection episodes}.

\begin{acknowledgements}
DMF, ACC and ADS thank the CNRS-PICS program for supporting our Brazilian-French collaboration and the present work. 
DMF acknowledges support from FAPESP (grant 2012/04916-7). ACC acknowledges support from CNPq (grant 307076/2012-1) and FAPESP (grant 2010/19029-0).
This work has made use of the computing facilities of the Laboratory of Astroinformatics (IAG/USP, NAT/Unicsul), whose purchase was made possible by the Brazilian agency FAPESP (grant 2009/54006-4) and the INCT-A. 
\end{acknowledgements}

\bibliographystyle{aa}
\bibliography{bib}

\begin{thebibliography}{30}
\expandafter\ifx\csname natexlab\endcsname\relax\def\natexlab#1{#1}\fi

\bibitem[{{Bjorkman}(1997)}]{bjo97}
{Bjorkman}, J.~E. 1997, in Lecture Notes in Physics, Berlin Springer Verlag,
  Vol. 497, Stellar Atmospheres: Theory and Observations, ed. J.~P. {De Greve},
  R.~{Blomme}, \& H.~{Hensberge}, 239

\bibitem[{{Bjorkman} \& {Carciofi}(2005)}]{bjo05}
{Bjorkman}, J.~E. \& {Carciofi}, A.~C. 2005, in Astronomical Society of the
  Pacific Conference Series, Vol. 337, The Nature and Evolution of Disks Around
  Hot Stars, ed. {R.~Ignace \& K.~G.~Gayley}, 75

\bibitem[{{Born} \& {Wolf}(1980)}]{bor80}
{Born}, M. \& {Wolf}, E. 1980, {Principles of Optics Electromagnetic Theory of
  Propagation, Interference and Diffraction of Light}

\bibitem[{{Carciofi}(2011)}]{car11}
{Carciofi}, A.~C. 2011, in Active OB Stars, Vol. 272, IAU Symposium, ed.
  C.~{Neiner}, G.~{Wade}, G.~{Meynet}, \& G.~{Peters}, 325--336

\bibitem[{{Carciofi} \& {Bjorkman}(2006)}]{car06}
{Carciofi}, A.~C. \& {Bjorkman}, J.~E. 2006, \apj, 639, 1081

\bibitem[{{Carciofi} \& {Bjorkman}(2008)}]{car08}
{Carciofi}, A.~C. \& {Bjorkman}, J.~E. 2008, \apj, 684, 1374

\bibitem[{{Carciofi} {et~al.}(2012){Carciofi}, {Bjorkman}, {Otero}, {Okazaki},
  {{\v S}tefl}, {Rivinius}, {Baade}, \& {Haubois}}]{car12}
{Carciofi}, A.~C., {Bjorkman}, J.~E., {Otero}, S.~A., {et~al.} 2012, \apjl,
  744, L15

\bibitem[{{Carciofi} {et~al.}(2009){Carciofi}, {Okazaki}, {Le Bouquin}, {{\v
  S}tefl}, {Rivinius}, {Baade}, {Bjorkman}, \& {Hummel}}]{car09}
{Carciofi}, A.~C., {Okazaki}, A.~T., {Le Bouquin}, J.-B., {et~al.} 2009, \aap,
  504, 915

\bibitem[{{Domiciano de Souza} {et~al.}(2004){Domiciano de Souza}, {Zorec},
  {Jankov}, {Vakili}, {Abe}, \& {Janot-Pacheco}}]{dom04}
{Domiciano de Souza}, A., {Zorec}, J., {Jankov}, S., {et~al.} 2004, \aap, 418,
  781

\bibitem[{{Faes} {et~al.}(2012){Faes}, {Carciofi}, {Rivinius}, {{\v S}tefl},
  {Baade}, \& {Domiciano de Souza}}]{fae12}
{Faes}, D.~M., {Carciofi}, A.~C., {Rivinius}, T., {et~al.} 2012, in
  Circumstellar Dynamics at High Resolution, Vol. 464, PASPC, ed. A.~C.
  {Carciofi} \& T.~{Rivinius}, 141

\bibitem[{{Hanuschik}(1995)}]{han95}
{Hanuschik}, R.~W. 1995, \aap, 295, 423

\bibitem[{{Harmanec}(1994)}]{har94}
{Harmanec}, P. 1994, in NATO ASIC Proc. 436: The Impact of Long-Term Monitoring
  on Variable Star Research: Astrophysics, ed. C.~{Sterken} \& M.~{de Groot},
  55

\bibitem[{{Harmanec}(2000)}]{har00}
{Harmanec}, P. 2000, in Astronomical Society of the Pacific Conference Series,
  Vol. 214, IAU Colloq. 175: The Be Phenomenon in Early-Type Stars, ed. M.~A.
  {Smith}, H.~F. {Henrichs}, \& J.~{Fabregat}, 13

\bibitem[{{Haubois} {et~al.}(2012){Haubois}, {Carciofi}, {Rivinius}, {Okazaki},
  \& {Bjorkman}}]{hau12}
{Haubois}, X., {Carciofi}, A.~C., {Rivinius}, T., {Okazaki}, A.~T., \&
  {Bjorkman}, J.~E. 2012, \apj, 756, 156

\bibitem[{{Jankov} {et~al.}(2001){Jankov}, {Vakili}, {Domiciano de Souza}, \&
  {Janot-Pacheco}}]{jan01}
{Jankov}, S., {Vakili}, F., {Domiciano de Souza}, Jr., A., \& {Janot-Pacheco},
  E. 2001, \aap, 377, 721

\bibitem[{{Jones} {et~al.}(2008){Jones}, {Sigut}, \& {Porter}}]{jon08}
{Jones}, C.~E., {Sigut}, T.~A.~A., \& {Porter}, J.~M. 2008, \mnras, 386, 1922

\bibitem[{{Kraus} {et~al.}(2012){Kraus}, {Monnier}, {Che}, {Schaefer},
  {Touhami}, {Gies}, {Aufdenberg}, {Baron}, {Thureau}, {ten Brummelaar},
  {McAlister}, {Turner}, {Sturmann}, \& {Sturmann}}]{kra11}
{Kraus}, S., {Monnier}, J.~D., {Che}, X., {et~al.} 2012, \apj, 744, 19

\bibitem[{{Lee} {et~al.}(1991){Lee}, {Osaki}, \& {Saio}}]{lee91}
{Lee}, U., {Osaki}, Y., \& {Saio}, H. 1991, \mnras, 250, 432

\bibitem[{{McGill} {et~al.}(2011){McGill}, {Sigut}, \& {Jones}}]{mcg11}
{McGill}, M.~A., {Sigut}, T.~A.~A., \& {Jones}, C.~E. 2011, \apj, 743, 111

\bibitem[{{Meilland} {et~al.}(2007){Meilland}, {Stee}, {Vannier}, {Millour},
  {Domiciano de Souza}, {Malbet}, {Martayan}, {Paresce}, {Petrov}, {Richichi},
  \& {Spang}}]{mei07}
{Meilland}, A., {Stee}, P., {Vannier}, M., {et~al.} 2007, \aap, 464, 59

\bibitem[{{Mourard} {et~al.}(2009){Mourard}, {Clausse}, {Marcotto}, {Perraut},
  {Tallon-Bosc}, {B{\'e}rio}, {Blazit}, {Bonneau}, {Bosio}, {Bresson},
  {Chesneau}, {Delaa}, {H{\'e}nault}, {Hughes}, {Lagarde}, {Merlin}, {Roussel},
  {Spang}, {Stee}, {Tallon}, {Antonelli}, {Foy}, {Kervella}, {Petrov},
  {Thiebaut}, {Vakili}, {McAlister}, {ten Brummelaar}, {Sturmann}, {Sturmann},
  {Turner}, {Farrington}, \& {Goldfinger}}]{mou09}
{Mourard}, D., {Clausse}, J.~M., {Marcotto}, A., {et~al.} 2009, \aap, 508, 1073

\bibitem[{{Okazaki}(2001)}]{oka01}
{Okazaki}, A.~T. 2001, \pasj, 53, 119

\bibitem[{{Petrov} {et~al.}(2007){Petrov}, {Malbet}, {Weigelt}, {Antonelli},
  {Beckmann}, {Bresson}, {Chelli}, {Dugu{\'e}}, {Duvert}, {Gennari},
  {Gl{\"u}ck}, {Kern}, {Lagarde}, {Le Coarer}, {Lisi}, {Millour}, {Perraut},
  {Puget}, {Rantakyr{\"o}}, {Robbe-Dubois}, {Roussel}, {Salinari}, {Tatulli},
  {Zins}, {Accardo}, {Acke}, {Agabi}, {Altariba}, {Arezki}, {Aristidi},
  {Baffa}, {Behrend}, {Bl{\"o}cker}, {Bonhomme}, {Busoni}, {Cassaing},
  {Clausse}, {Colin}, {Connot}, {Delboulb{\'e}}, {Domiciano de Souza},
  {Driebe}, {Feautrier}, {Ferruzzi}, {Forveille}, {Fossat}, {Foy},
  {Fraix-Burnet}, {Gallardo}, {Giani}, {Gil}, {Glentzlin}, {Heiden},
  {Heininger}, {Hernandez Utrera}, {Hofmann}, {Kamm}, {Kiekebusch}, {Kraus},
  {Le Contel}, {Le Contel}, {Lesourd}, {Lopez}, {Lopez}, {Magnard}, {Marconi},
  {Mars}, {Martinot-Lagarde}, {Mathias}, {M{\`e}ge}, {Monin}, {Mouillet},
  {Mourard}, {Nussbaum}, {Ohnaka}, {Pacheco}, {Perrier}, {Rabbia}, {Rebattu},
  {Reynaud}, {Richichi}, {Robini}, {Sacchettini}, {Schertl}, {Sch{\"o}ller},
  {Solscheid}, {Spang}, {Stee}, {Stefanini}, {Tallon}, {Tallon-Bosc}, {Tasso},
  {Testi}, {Vakili}, {von der L{\"u}he}, {Valtier}, {Vannier}, \&
  {Ventura}}]{pet07}
{Petrov}, R.~G., {Malbet}, F., {Weigelt}, G., {et~al.} 2007, \aap, 464, 1

\bibitem[{{Porter} \& {Rivinius}(2003)}]{por03}
{Porter}, J.~M. \& {Rivinius}, T. 2003, \pasp, 115, 1153

\bibitem[{{Quirrenbach} {et~al.}(1997){Quirrenbach}, {Bjorkman}, {Bjorkman},
  {Hummel}, {Buscher}, {Armstrong}, {Mozurkewich}, {Elias}, \&
  {Babler}}]{qui97}
{Quirrenbach}, A., {Bjorkman}, K.~S., {Bjorkman}, J.~E., {et~al.} 1997, \apj,
  479, 477

\bibitem[{{Rivinius} {et~al.}(1999){Rivinius}, {{\v S}tefl}, \&
  {Baade}}]{riv99}
{Rivinius}, T., {{\v S}tefl}, S., \& {Baade}, D. 1999, \aap, 348, 831

\bibitem[{{Stee}(1996)}]{ste96}
{Stee}, P. 1996, \aap, 311, 945

\bibitem[{{{\v S}tefl} {et~al.}(2009){{\v S}tefl}, {Rivinius}, {Carciofi}, {Le
  Bouquin}, {Baade}, {Bjorkman}, {Hesselbach}, {Hummel}, {Okazaki}, {Pollmann},
  {Rantakyr{\"o}}, \& {Wisniewski}}]{ste09}
{{\v S}tefl}, S., {Rivinius}, T., {Carciofi}, A.~C., {et~al.} 2009, \aap, 504,
  929

\bibitem[{{von Zeipel}(1924)}]{von24}
{von Zeipel}, H. 1924, \mnras, 84, 665

\bibitem[{{Wheelwright} {et~al.}(2012){Wheelwright}, {Bjorkman}, {Oudmaijer},
  {Carciofi}, {Bjorkman}, \& {Porter}}]{whe12}
{Wheelwright}, H.~E., {Bjorkman}, J.~E., {Oudmaijer}, R.~D., {et~al.} 2012,
  \mnras, 423, L11

\end{thebibliography}
 
\end{document}